\documentclass[10pt,a4paper,twocolumn]{article}

\usepackage[a4paper,margin=1.7cm]{geometry}
\usepackage[T1]{fontenc}
\usepackage[utf8]{inputenc}
\usepackage{lmodern}
\usepackage{microtype}

\usepackage{amsmath,amssymb,amsthm}
\usepackage{mathtools}

\usepackage{booktabs}
\usepackage{multirow}
\usepackage{array}
\usepackage{tabularx}

\usepackage{graphicx}
\usepackage{subcaption}
\usepackage{tikz}
\usetikzlibrary{arrows.meta,positioning,fit,calc,shapes.geometric,shapes.multipart,decorations.pathreplacing,backgrounds}
\usepackage{pgfplots}
\pgfplotsset{compat=1.18}
\usepgfplotslibrary{groupplots}

\usepackage[colorlinks=true,linkcolor=blue,urlcolor=blue,citecolor=blue]{hyperref}
\usepackage{doi}  
\usepackage{cleveref}

\usepackage{chemformula}

\usepackage[numbers,sort&compress]{natbib}

\usepackage{listings}
\usepackage{xcolor}
\usepackage[normalem]{ulem}   
\definecolor{revgreen}{rgb}{0,0.55,0}
\usepackage{enumitem}
\setlist{nosep}

\usepackage{seqsplit}
\usepackage{dblfloatfix}
\usepackage{sectsty}
\allsectionsfont{\bfseries\raggedright}
\newcommand{\smi}[1]{\texttt{\seqsplit{#1}}}

\newcommand{\xvr}{\textsc{XTP}}
\newcommand{\badt}{G_{\mathrm{ADT}}}
\newcommand{\bper}[1]{G(#1)}
\newcommand{\adt}{\textsc{ADT}}
\newcommand{\ikt}{\textsc{IKT}}

\newcommand{\angstrom}{\text{\AA}}

\title{\textbf{Atomic Design Transformer: \\
  Scaffold-Conditioned 3D Molecule Generation with xTB-Verified Reinforcement Learning}}
\author{Takao Kotani\\
  \\
  AI for Materials Innovation Hub, Institute for Advanced Study, Kyoto University,\\
  Yoshida, Sakyo, Kyoto 606-8501, JAPAN,\\
  Center for Spintronics Research Network, Osaka University, Toyonaka 560-8531, Japan,\\
  Division of Electrical, Electronic and Information Engineering, Osaka University,\\
  Suita 565-0871, Japan\\
  \texttt{takaokotani@gmail.com}
}

\date{\today}

\begin{document}
\twocolumn[%
\begin{@twocolumnfalse}
\maketitle

\begin{abstract}
  We present an autoregressive 3D-molecule generator with SE(3)-invariant tokenization,
  the Atomic Design Transformer (\adt{}). \adt{} places atoms one
  at a time, autoregressively. SE(3) invariance is achieved by
  tokenization: each new atom's position is encoded in the local
  coordinate frame of a previously placed atom. The backbone is a plain
  causal transformer. The token stream fully specifies a 3D structure
  together with its chemical-bond graph $\badt$, without any bond-order
  assignment. The model emits heavy-atom skeletons; hydrogens are added by
  separate learned models before the xTB relaxation. To score generated molecules we introduce the xTB
  topology-preservation rate (\xvr{}): the fraction of molecules for
  which an xTB~GFN2 relaxation preserves $\badt$ specified by the token stream. 
  For \xvr{}-accepted molecules we also report the
  relaxation energy and the root mean square of the atomic displacement (RMSD). We evaluate
  two \adt{} models. The first is \adt{} pretrained on the GEOM-Drugs
  $\le\!30$-heavy-atom dataset; we benchmark scaffold-conditioned 3D
  generation across seven drug-like scaffolds from the model. It
  reaches an \xvr{} of ${\sim}55\%$ and a valid-molecule yield
  $N^{\mathrm{gen}}/N$ of ${\sim}53\%$, where $N^{\mathrm{gen}}/N$ is the
  fraction of samples that are distinct, topology-preserving, and
  RDKit-readable. The second model continues from the first by
  reinforcement learning against the verifiable xTB reward (RLVR), using
  no external molecules. RLVR raises \xvr{} to ${\sim}98\%$ and
  $N^{\mathrm{gen}}/N$ to ${\sim}95\%$, while approximately preserving the
  GEOM-Drugs size and composition distributions. Finally, we present 
  an Inverse-Kinematics Transformer that recovers \xvr{} for large molecules, where discretization error accumulates.
  \adt{} thus enables direct 3D generation.
\end{abstract}
\bigskip
\end{@twocolumnfalse}
]

\section{Introduction}
\label{sec:intro}

The \emph{in silico} design of 3D molecules that satisfy prescribed functional
criteria is a central open problem in generative modeling for chemistry. The recent abundance of
large molecular databases such as GEOM-Drugs~\citep{geom2022} and the rapid progress of deep generative models have
made data-driven molecular design both attractive and tractable, yet fully
automated 3D generation 
remains an open problem.

Direct 3D generators are attractive because they learn geometry and chemistry
jointly, enabling native conditioning on spatial context such as binding
pockets. Current direct 3D methods fall into two families: diffusion-based/flow-matching-based
generators, and autoregressive (AR) models that build molecules atom-by-atom.
The recent AR transformers (Quetzal~\citep{quetzal2025}, InertialAR~\citep{inertialar2025})
are still hybrid, emitting the atomic position through a separate
generator rather than from the transformer itself.

In this paper we present the Atomic Design Transformer (\adt{}), 
which generates 3D molecules atom-by-atom by encoding each atomic position directly as a discrete token 
in a local coordinate frame anchored on a previously placed atom. 
The transformer emits the geometry itself, with no separate position generator. 
Combined with reinforcement learning (RL), \adt{} can generate a wider range of molecules 
not by learning a wider range of GEOM-Drugs-like data 
but by learning the rules defined by the xTB GFN2 method~\citep{xtb2019}.

A caution before we proceed. Because \adt{} learns xTB's rules rather than a data distribution, it
inherits xTB's approximations (in addition, we have room to improve our RL).
A molecule can be valid under both xTB and RDKit yet still look
unrealistic to a chemist. Our analysis finds ${\sim}3\%$ of generated molecules may be problematic, such
as charge-separated carbanion--cation pairs or strained rings (\S\ref{sec:drugs50}).

\paragraph{Scaffold-conditional generation}
As a starting point, we focus on scaffold-conditional generation. Given
a scaffold, such as benzene, the model must generate a complete, valid
3D molecule that contains it, keeping the scaffold's atoms fixed and
generating the remaining atoms around them. This is a basic form of
controllable 3D generation. Quetzal~\citep{quetzal2025} highlighted it
as a natural capability of autoregressive generation, since the scaffold
can be supplied as a fixed token prefix, with no architectural change or
scaffold-specific retraining. It demonstrates this on three scaffolds (benzene,
1,2,4-triazole, and thiophene) but states that ``these are qualitative
results; we defer quantitative evaluation and comparison to future
work.'' 
Providing such a benchmark with \adt{} is a main goal of this paper.

To check the 3D quality of generated molecular structures,
we devise a new metric, the xTB topology-preservation rate
(\xvr{}), which checks whether the structure preserves its
heavy-atom bond topology (read from the 3D geometry; \S\ref{sec:xvr})
after relaxation with xTB GFN2~\citep{xtb2019}.
The metric is practical because xTB GFN2 is the \emph{de facto}
workhorse for geometry optimization of drug-like organic molecules;
GEOM-Drugs itself was generated with it, so the reference data and the
metric rest on the same physics. For \xvr{}-accepted molecules we further report the relaxation energy (strain) and the root mean square of the atomic displacement (RMSD) on relaxation.

\paragraph{Key idea: SE(3)-invariant tokenization with a vanilla transformer}
Most prior 3D-molecule generators have assumed that SE(3) invariance
must be enforced inside the network through equivariant
message passing or diffusion (E(3)-equivariant GNNs, equivariant
diffusion, and so on). The community has invested years along this line. 
Instead of such approaches, we put the invariance into the
input: every atomic position is encoded in a local coordinate frame
anchored on a previously placed atom. The combination of discrete
tokenization + local-frame coordinates + standard transformer is a
notably simple design.
This design is also advantageous because we can use the standard
transformer techniques such as KV caching, batched inference, prefix conditioning, fine-tuning, 
and distributed training.

\paragraph{Our contributions.}

\begin{enumerate}
\item \textbf{Fully discrete AR architecture, no
        equivariant layers.} We describe a molecule as a stream of ADT tokens. 
        The \adt{} grammar is simple. Atom species, parent atoms, and positions are all tokenized with fixed-size
        vocabularies. SE(3) invariance is ensured  
        by construction since we place atoms in a local coordinate frame of the parent atom.
        We process the token streams with a causal transformer.
\item \textbf{\xvr{} (xTB-topology-preservation).}
        We introduce \xvr{}, which asks whether the heavy-atom bond graph $\badt$ declared by the token stream is preserved under xTB GFN2 relaxation.
        We validate models with a physically grounded evaluation (\xvr{}, xTB strain, and RMSD).
\item \textbf{Data-free RLVR for 3D generation.} 
        We evaluate two models. Using the pretrained model
        as a prior, we refine it by reinforcement learning from the verifiable
        xTB reward (RLVR) with no external data (self-improvement against a verifiable physical reward).
\item \textbf{Scaffold-conditional generation benchmark.}
        Seven drug-like scaffolds (benzene, pyridine, pyrimidine, pyrazine,
        furan, thiophene, cyclohexane) $\times$ $N{=}10{,}000$. 
        These scaffolds are supplied as a prefix.
  \item \textbf{Inverse-Kinematics Transformer for the scaling problem.}
        To avoid error accumulation for long molecules, we introduce IKT, a bidirectional transformer that corrects the ADT output.
\end{enumerate}

\begin{figure*}[t!]
\centering
\includegraphics[width=\linewidth]{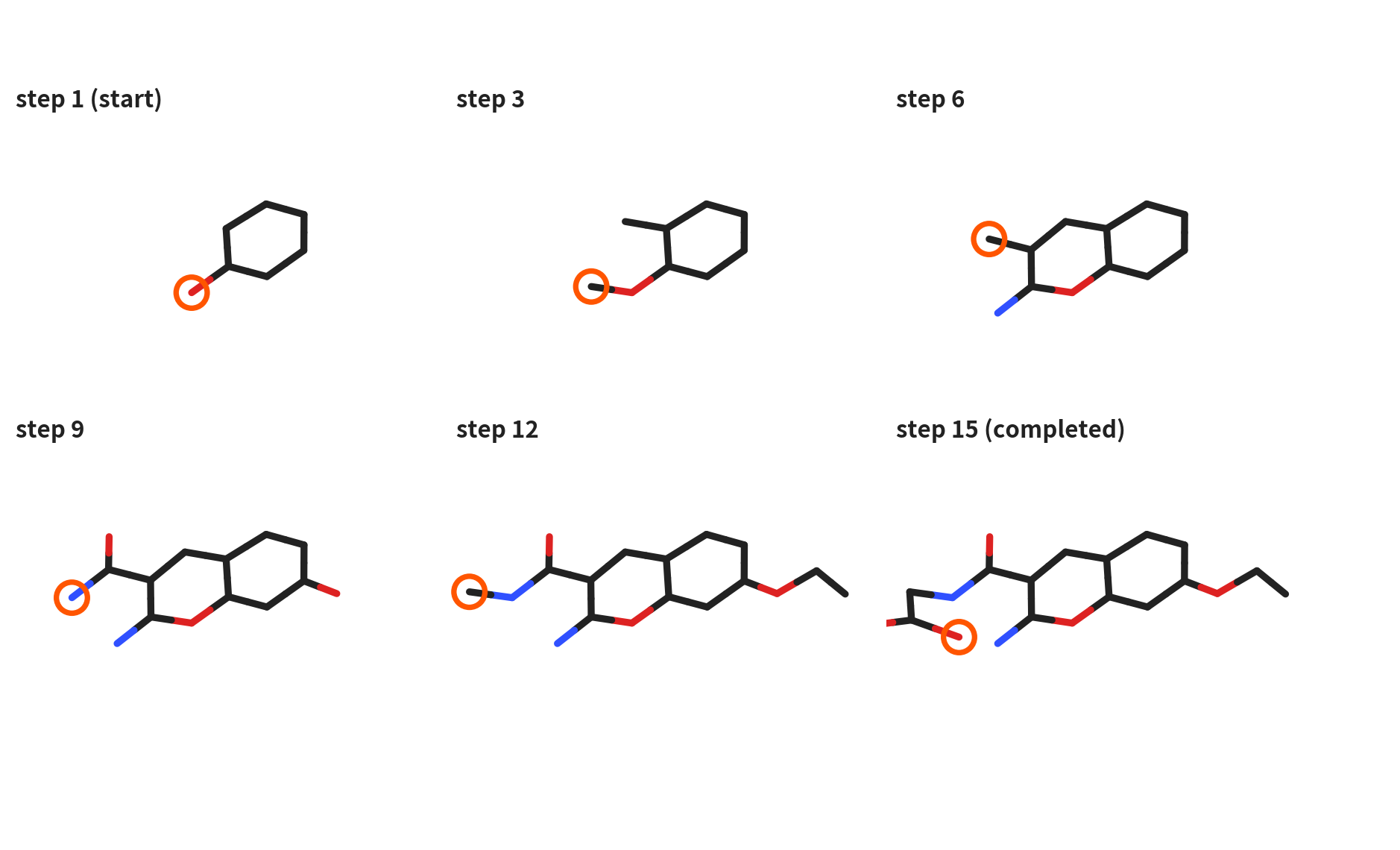}
\caption{\textbf{Scaffold-conditional generation, one atom at a time.}
  \adt{} is prompted with a fixed seed scaffold (here the benzene
  ring) and then autoregressively places one heavy atom per step; the
  orange circle marks the atom added at that step (selected steps
  shown: 1, 3, 6, 9, 12, and the final molecule at step~15). The model
  decides when to stop, terminating generation on its own.
  See \url{https://tkotani.github.io/ADT/benzene.html}.}
\label{fig:scaffold-growth}
\end{figure*}

Figure~\ref{fig:scaffold-growth} illustrates this scaffold-conditional,
atom-by-atom construction on a benzene seed.
Code, model weights, and the \xvr{} evaluation tool are available at
\url{https://github.com/tkotani/ADT}; see \S\ref{sec:repro}.

\section{Related Work}
\label{sec:related}

On GEOM-Drugs, recent diffusion and flow-matching
methods~\citep{midi2023,eqgatdiff2024,semlaflow2025,georcg2024,canonflow2026,lenses2026} report almost perfect
($\sim$$100\%$) generation quality. These values are not directly comparable with XTP: they use different
molecular representations, validity definitions and relaxation protocols. With \adt{}, the unconditional (random-Btriple, \S\ref{sec:drugs30}) end-to-end yield of valid molecules reaches
$96.7\%$ (Table~\ref{tab:drugs30}) under our XTP criterion. 
However, such numbers can be reduced easily if we reduce the allowed maximum of the relaxation energy.

\paragraph{Diffusion {and flow-matching} 3D generators.}
EDM~\citep{edm2022} founded this family: an E(3)-equivariant diffusion
over continuous atom coordinates and types, trained unconditionally on
QM9 and GEOM-Drugs. GeoLDM~\citep{geoldm2023} runs the same diffusion in
an equivariant latent space for scalability, while
MiDi~\citep{midi2023} co-generates the 2D bond graph together with
the coordinates (discrete graph diffusion $+$ continuous geometry).
GCDM~\citep{gcdm2024} adds geometry-complete message passing for larger
drug-like molecules, and NExT-Mol~\citep{nextmol2025} couples a 1D SMILES
language model with a 3D diffusion head.
TargetDiff~\citep{targetdiff2023} and DiffSBDD~\citep{diffsbdd2024} carry
the same equivariant machinery to pocket-conditioned
structure-based drug design.

{Flow-matching models replace the diffusion with a learned
continuous flow that straightens the probability path; SemlaFlow~\citep{semlaflow2025}
and the representation-conditioned GeoRCG~\citep{georcg2024} and LENSEs~\citep{lenses2026} follow this
line, while CanonFlow~\citep{canonflow2026} is based on its own canonicalization.}

\paragraph{Autoregressive 3D generators.}
The autoregressive setting closest to \adt{} is occupied by two recent
families. The hybrid transformers Quetzal~\citep{quetzal2025} and
InertialAR~\citep{inertialar2025} emit a per-atom latent and draw the
discrete atom type from a softmax, but produce the continuous
position from a separate downstream generator (a diffusion MLP and a
Gaussian head, respectively); the transformer never emits the position
itself. The discrete-token language models
Geo2Seq~\citep{geo2seq} and Mol-StrucTok~\citep{molstructok} are fully
tokenized. Earlier autoregressive
generators include G-SchNet~\citep{gschnet2019} (atom-by-atom placement
with SE(3)-invariant point-cloud features), Symphony~\citep{symphony2024}
(E(3)-equivariant message passing with spherical-harmonic directions),
Frag2Seq~\citep{frag2seq}, Flam-Shepherd \&
Aspuru-Guzik~\citep{flamshepherd2023}, and G-SphereNet~\citep{gspherenet2022}.

\paragraph{Reinforcement learning.}
MolGym~\citep{molgym2020} builds 3D molecules from scratch by RL against
a verifiable quantum-mechanical energy reward; closer to our setting,
RLPF~\citep{rlpf2025} fine-tunes a pretrained equivariant diffusion
model with PPO against a physical (force-field) stability reward. \adt{}'s
use of RL is complementary: it trains a discrete autoregressive
transformer, and the data-free reward is \xvr{} with the
xTB~\citep{xtb2019} GFN2 relaxation energy.

\paragraph{Scaffold-conditioned generation.}
Fragment-based 2D approaches (DrugEx, SCAAR); 3D linker design
(DeLinker~\citep{delinker2020}, DiffLinker~\citep{difflinker2024}); 3D
scaffold hopping (DiffHopp~\citep{diffhopp2023}); shape-conditioned
(SQUID~\citep{squid2023}). DiffDec~\citep{diffdec2024} is a pocket-aware diffusion model for
R-group decoration on a fixed Bemis--Murcko-like scaffold; the task
is distinct from ours (pocket-conditioned R-group restoration). To the best of our knowledge, no prior work reports
quantitative $N{=}10{,}000$ scaffold-conditional 3D-generation
benchmarks with quantum-mechanical geometry validation across
multiple scaffolds.

\section{Method}
\label{sec:method}

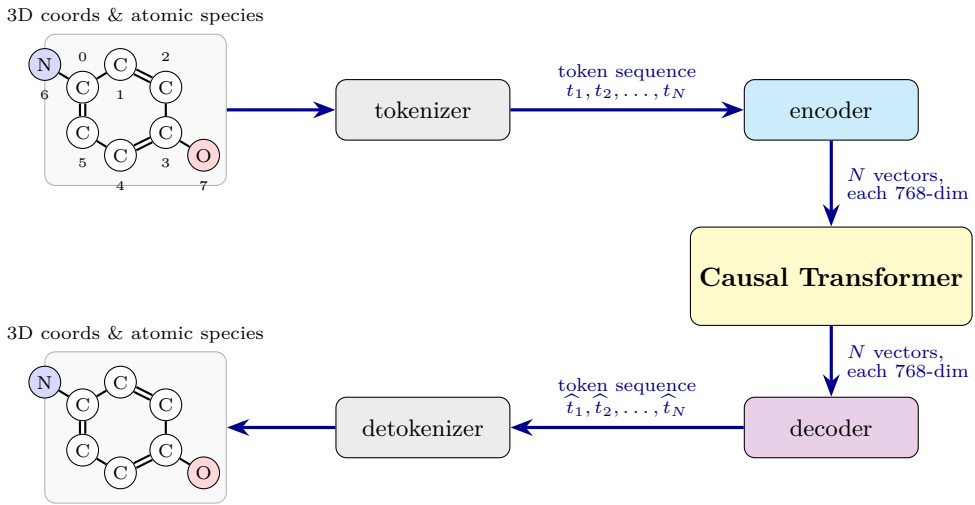
\begin{figure*}[t!]
  \centering
  \begin{tikzpicture}[
      font=\small,
      atom/.style={circle, draw, minimum size=0.42cm, fill=white, inner sep=0pt,
          font=\scriptsize},
      bond/.style={thick},
      dbond/.style={thick, double, double distance=1.2pt},
      mol3d/.style={rectangle, rounded corners=4pt, draw=gray!60, fill=gray!5,
          inner sep=5pt, minimum width=2.4cm, minimum height=2.0cm,
          align=center},
      op/.style={draw, rectangle, rounded corners=4pt, minimum height=0.8cm,
          minimum width=2.3cm, align=center, font=\small},
      enc/.style={op, fill=cyan!18},
      dec/.style={op, fill=violet!18},
      tokenizer/.style={op, fill=gray!15},
      tf/.style={draw, rectangle, rounded corners=4pt, minimum height=1.3cm,
          minimum width=3.0cm, fill=yellow!25, align=center,
          font=\normalsize\bfseries},
      arr/.style={-Stealth, very thick, color=blue!55!black},
    ]

    \node[mol3d] (inmol) at (0, 0) {};
    \node[anchor=south, font=\scriptsize] at (0, 1.00) {3D coords \& atomic species};
    \node[atom, label={[font=\tiny]above:0}]  (p0) at (-0.7, 0.30) {C};
    \node[atom, label={[font=\tiny]below:1}]  (p1) at (-0.2, 0.60) {C};
    \node[atom, label={[font=\tiny]above:2}]  (p2) at ( 0.4, 0.30) {C};
    \node[atom, label={[font=\tiny]below:3}]  (p3) at ( 0.4,-0.30) {C};
    \node[atom, label={[font=\tiny]below:4}]  (p4) at (-0.2,-0.60) {C};
    \node[atom, label={[font=\tiny]below:5}]  (p5) at (-0.7,-0.30) {C};
    \node[atom, fill=blue!15, label={[font=\tiny]below:6}] (pN) at (-1.2, 0.60) {N};
    \node[atom, fill=red!15,  label={[font=\tiny]below:7}] (pO) at ( 0.9,-0.60) {O};
    \draw[bond]  (p0)--(p1); \draw[dbond] (p1)--(p2);
    \draw[bond]  (p2)--(p3); \draw[dbond] (p3)--(p4);
    \draw[bond]  (p4)--(p5); \draw[dbond] (p5)--(p0);
    \draw[bond]  (p0)--(pN); \draw[bond] (p3)--(pO);

    \node[tokenizer] (tok) at (3.8, 0) {tokenizer};
    \node[enc]       (e)   at (9.2, 0) {encoder};

    \draw[arr] (inmol.east) -- (tok.west);
    \draw[arr] (tok.east) -- (e.west)
    node[midway, above, font=\scriptsize, align=center]
    {token sequence\\[-1pt]\scriptsize$t_1,t_2,\dots,t_N$};

    \node[tf]  (trans) at (9.2, -2.2) {Causal Transformer};
    \node[dec] (d)     at (9.2, -4.2) {decoder};

    \draw[arr] (e.south) -- (trans.north)
    node[midway, right=2pt, font=\scriptsize, align=left]
    {$N$ vectors,\\[-1pt]each 768-dim};
    \draw[arr] (trans.south) -- (d.north)
    node[midway, right=2pt, font=\scriptsize, align=left]
    {$N$ vectors,\\[-1pt]each 768-dim};

    \node[tokenizer] (detok) at (3.8, -4.2) {detokenizer};
    \node[mol3d] (outmol) at (0, -4.2) {};
    \node[anchor=south, font=\scriptsize] at (0, -3.20) {3D coords \& atomic species};
    \node[atom] (q0) at (-0.7, -3.90) {C};
    \node[atom] (q1) at (-0.2, -3.60) {C};
    \node[atom] (q2) at ( 0.4, -3.90) {C};
    \node[atom] (q3) at ( 0.4, -4.50) {C};
    \node[atom] (q4) at (-0.2, -4.80) {C};
    \node[atom] (q5) at (-0.7, -4.50) {C};
    \node[atom, fill=blue!15] (qN) at (-1.2, -3.60) {N};
    \node[atom, fill=red!15]  (qO) at ( 0.9, -4.80) {O};
    \draw[bond]  (q0)--(q1); \draw[dbond] (q1)--(q2);
    \draw[bond]  (q2)--(q3); \draw[dbond] (q3)--(q4);
    \draw[bond]  (q4)--(q5); \draw[dbond] (q5)--(q0);
    \draw[bond]  (q0)--(qN); \draw[bond] (q3)--(qO);

    \draw[arr] (d.west) -- (detok.east)
    node[midway, above, font=\scriptsize, align=center]
    {token sequence\\[-1pt]\scriptsize$\widehat{t}_1,\widehat{t}_2,\dots,\widehat{t}_N$};
    \draw[arr] (detok.west) -- (outmol.east);

  \end{tikzpicture}
  \caption{\adt{} pipeline as a two-row horizontal data flow.
    \emph{Top row (left $\to$ right):} 3D coordinates with atomic species
    $\to$ tokenizer $\to$ token sequence $(t_1,\dots,t_N)$ $\to$ encoder
    (each token $t_i$ is embedded into its own 768-dim hidden vector, so
    the transformer sees a sequence of length $N$, not $N/7$).
    \emph{Pullback (right):} hidden vectors flow down through the causal
    transformer into the decoder, which emits next-token logits.
    \emph{Bottom row (right $\to$ left):} sampled tokens
    $\widehat{t}_1,\dots,\widehat{t}_N$ $\to$ detokenizer $\to$ output 3D
    coordinates. A concrete example of the token sequence produced by this
    pipeline is given in Table~\ref{tab:toksample}.}
  \label{fig:pipeline}
\end{figure*}

The essential idea of \adt{} is to tokenize a 3D atomic configuration
into a sequence of discrete tokens that describe how each atom is placed
relative to previously placed atoms, train an AR transformer on
these sequences, and reconstruct 3D atomic coordinates from the model's sampled
tokens at inference time. Figure~\ref{fig:pipeline} illustrates this
encode--decode round-trip.

A crucial ingredient of this formulation is that every placed atom
  carries its own local coordinate frame. When a new atom is added, its position
is not expressed in any absolute (world) coordinate system, but as a relative
displacement in the local frame of its parent atom, namely a distance to the
parent and a direction in the parent's frame. As soon as the new atom is
placed, its own local frame is constructed from the bond from the atom to the
parent and the bond from the parent to the parent's parent
(\S\ref{sec:localframe}). This per-atom local-frame construction makes the
entire sequence manifestly invariant under global rotations and translations of
the molecule, without any equivariant architectural machinery; the model itself
is a plain causal transformer over discrete tokens. The discretization of the
relative distance and direction is described in \S\ref{sec:tokenization}.

\begin{table}[htb]
  \centering
  \small
  \setlength{\tabcolsep}{5pt}
  \begin{tabular}{c@{~~}c|l@{~~}r@{~~}r@{~~}r@{~~}r@{~~}r@{~~}r}
    \toprule
    step      & AtomID & action         & offset & $Z$          & $r_b$ & $h_0$        & $h_1$         & $h_2$ \\
    \midrule
    $0^{\,*}$ & A0     & \texttt{INIT}  & ---    & 6            & ---   & ---          & ---           & ---   \\
    $1^{\,*}$ & A1     & \texttt{CHAIN} & ---    & 6            & 96    & ---          & ---           & ---   \\
    $2^{\,*}$ & A2     & \texttt{ANGLE} & ---   & 6            & 98    & $7^\ddagger$ & $15^\ddagger$ & ---   \\
    3         & A3     & \texttt{ADD}   & $-1$   & 6            & 96    & 4            & 0             & 6     \\
    4         & A4     & \texttt{ADD}   & $-4$   & 6            & 97    & 4            & 15            & 9     \\
    5         & A5     & \texttt{ADD}   & $-5$   & 6            & 97    & 4            & 0             & 6     \\
    6         & ---    & \texttt{LINK}  & $-6$   & $-3^\dagger$ & 98    & 4            & 0             & 9     \\
    7         & A6     & \texttt{ADD}   & $-6$   & 7            & 98    & 2            & 15            & 13    \\
    8         & A7     & \texttt{ADD}   & $-1$   & 8            & 93    & 6            & 0             & 3     \\
    9         & ---    & \texttt{END}   &        &              &       &              &               &       \\
    \bottomrule
  \end{tabular}
  \caption{A concrete token sequence produced by our tokenizer for
    4-aminophenol (8 heavy atoms: aromatic ring + \ch{NH2} +
    \ch{OH}). Each of the first $N_{\mathrm{step}}-1$ rows carries 7
    tokens (action,\,offset,\,$Z$,\,$r_b$,\,$h_0$,\,
    $h_1$,\,$h_2$), followed by a single \texttt{END} token, giving
    $N_{\mathrm{step}}=10$ and $N=7(N_{\mathrm{step}}-1)+1=64$ total tokens
    (Eq.~\ref{eq:Ntokens}). AtomID labels
    atoms in placement order (A0\dots A7); offset points back by
    that many placed atoms, e.g.\ at step~4, offset~$-4$ means the parent
    is the atom placed 4 steps earlier (A0). The bootstrap triple (Btriple; steps
    0--2, marked $^{*}$) is a given prefix excluded from the training loss;
    targets start at step~3. Subsequent \texttt{ADD} tokens extend the
    aromatic ring, \texttt{LINK} closes it (between A0 and A3), and the
    final two \texttt{ADD}s place the amino nitrogen (A6) and the hydroxyl
    oxygen (A7).
    $^{\dagger}$For \texttt{LINK}, $Z$ is reused as a second offset
    (the closure partner).
    $^{\ddagger}$For \texttt{ANGLE}, $h_0,h_1$ are reused to hold tree-local
    angle bins $\theta_c,\theta_f$.
    $r_b$: log-spaced distance bin (Appendix~\ref{app:rbin}).
    $(h_0,h_1,h_2)$: HEALPix-encoded direction (Appendix~\ref{app:hpix}).}
  \label{tab:toksample}
\end{table}

\subsection{Tree structure and local coordinate frame}
\label{sec:localframe}

\paragraph{Atoms form a tree, with \texttt{LINK} providing the cycles.} Excluding the \texttt{LINK} action, each placement step
introduces exactly one new atom with a single, explicitly-pointed parent
(\S\ref{sec:offset}), so the sequence of placements
\texttt{INIT}$\to$\texttt{CHAIN}$\to$\texttt{ANGLE}$\to$\texttt{ADD}$\to\cdots$
builds a rooted tree $T$ rooted at the \texttt{INIT} atom. Every non-root atom
$A_n$ therefore has a unique chain of ancestors $A_{\pi(n)} \to A_{\pi^2(n)}
  \to \cdots \to A_0$ up to the root. A \texttt{LINK} action does not introduce a
new atom {but} draws an extra edge between two already-placed atoms, turning the
tree into a molecular graph that can contain rings. This tree structure over
atom additions is what enables a tractable ancestor-chain based construction of
local frames. The atom-placement dependencies form a directed acyclic graph, whereas the molecular graph
obtained after adding the \texttt{LINK} edges may contain cycles.

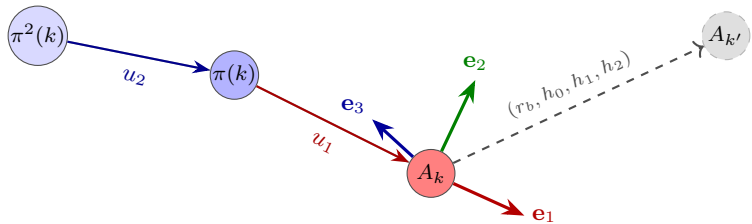
\begin{figure*}[t!]
\centering
\begin{tikzpicture}[scale=1.3,
    atom/.style={circle, draw=black!70, fill=#1, minimum size=18pt,
                 inner sep=0pt, font=\footnotesize},
    bond/.style={-, thick, gray!70},
    uvec/.style={-Stealth, thick, color=#1},
    axis/.style={-Stealth, very thick, color=#1},
]
\node[atom=blue!15]   (G) at (-2.6, 1.4) {$\pi^{2}(k)$};
\node[atom=blue!30]   (P) at (-0.6, 1.0) {$\pi(k)$};
\node[atom=red!50]    (K) at ( 1.4, 0.0) {$A_k$};

\draw[bond] (G) -- (P);
\draw[bond] (P) -- (K);

\draw[uvec=blue!55!black] (G) -- (P)
    node[midway, below=2pt, sloped, font=\small] {$u_2$};
\draw[uvec=red!65!black]  (P) -- (K)
    node[midway, below=2pt, sloped, font=\small] {$u_1$};

\draw[axis=red!70!black] (K) -- ++(0.95, -0.43)
    node[right=-1pt, font=\small\itshape] {$\mathbf{e}_1$};
\draw[axis=green!50!black] (K) -- ++(0.45, 0.95)
    node[above, font=\small\itshape] {$\mathbf{e}_2$};
\draw[axis=blue!60!black] (K) -- ++(-0.6, 0.55)
    node[above left=-2pt, font=\small\itshape] {$\mathbf{e}_3$};

\node[atom=gray!25, dashed, draw=gray!60]  (C) at (4.4, 1.4) {$A_{k'}$};
\draw[->, dashed, thick, gray!60!black] (K) -- (C)
    node[midway, above=1pt, sloped, font=\scriptsize]
    {$(r_b, h_0, h_1, h_2)$};

\end{tikzpicture}
\caption{Local coordinate frame $F_k$ at atom $A_k$.
  Two ancestor-chain edges
  $u_1 = p_k - p_{\pi(k)}$ (parent$\to A_k$) and
  $u_2 = p_{\pi(k)} - p_{\pi^2(k)}$ (grandparent$\to$parent) define an
  orthonormal triad
  $\{\mathbf{e}_1,\mathbf{e}_2,\mathbf{e}_3\}$ via Gram--Schmidt
  (Eq.~\ref{eq:gs-frame}): $\mathbf{e}_1$ is the unit vector along
  $u_1$, $\mathbf{e}_2$ is the residual of $u_2$ orthogonalized
  against $\mathbf{e}_1$, and $\mathbf{e}_3$ closes the right-handed
  frame. When the next atom $A_{k'}$ is added, the offset slot
  selects $A_k$ as its parent, and $A_{k'}$'s position is expressed
  in $F_k$ as the discrete tuple $(r_b, h_0, h_1, h_2)$
  ($r_b$: log-spaced distance bin; $(h_0,h_1,h_2)$: HEALPix-encoded
  unit direction). }
\label{fig:localframe}
\end{figure*}

\paragraph{Local frame stored at each placed atom.}
When a non-bootstrap atom $A_k$ is placed, we construct and store at
$A_k$ an orthonormal local frame $F_k \in SO(3)$ built from two consecutive
ancestor-chain edges:
\begin{equation}
  u_1 \;=\; p_k - p_{\pi(k)},
  \qquad
  u_2 \;=\; p_{\pi(k)} - p_{\pi^2(k)}.
  \label{eq:uvec}
\end{equation}
The frame axes are defined by a Gram--Schmidt procedure with $u_1$ as the
primary direction:
\begin{equation}
\begin{aligned}
  \mathbf{e}_1 &= \frac{u_1}{\|u_1\|}, \\
  \mathbf{e}_2 &= \frac{u_2 - (u_2\!\cdot\!\mathbf{e}_1)\mathbf{e}_1}
                      {\|u_2 - (u_2\!\cdot\!\mathbf{e}_1)\mathbf{e}_1\|}, \\
  \mathbf{e}_3 &= \mathbf{e}_1 \times \mathbf{e}_2,
  \qquad
  F_k = [\mathbf{e}_1\,\mathbf{e}_2\,\mathbf{e}_3]^{\top}.
\end{aligned}
  \label{eq:gs-frame}
\end{equation}
Thus $\mathbf{e}_1$ lies along the incoming bond (parent$\to A_k$) and
$\mathbf{e}_2$ fixes the plane with the previous ancestor edge, so the
triple $(r,\hat{u})$ of a subsequent child captures bond length, bond
angle, and dihedral angle in the familiar internal-coordinate sense
(Figure~\ref{fig:localframe}).

\paragraph{Local frame of the bootstrap atoms.}
Equation~\eqref{eq:uvec} requires two consecutive ancestor edges, which only
exist from $k\!\geq\!3$ onward: the bootstrap triple (Btriple) $A_0,A_1,A_2$ (actions
\texttt{INIT}, \texttt{CHAIN}, \texttt{ANGLE}) has an incomplete ancestor chain
($A_0$ has no parent; $A_1$ has only $A_0$; $A_2$ has only $A_1,A_0$). These
three atoms are therefore treated together: their positions jointly determine a
single bootstrap frame $F_{\mathrm{boot}}$ via Gram--Schmidt on
$u_1\!=\!p_2\!-\!p_1$ and $u_2\!=\!p_1\!-\!p_0$, and we set $F_0 \!=\! F_1
  \!=\! F_2 \!=\! F_{\mathrm{boot}}$ (a common local coordinate system shared by
all three bootstrap atoms). The first subsequent atom $A_3$ then uses
$F_2=F_{\mathrm{boot}}$ as its parent's frame in Eq.~\eqref{eq:localplace}, and
ordinary Eq.~\eqref{eq:uvec} takes over from $k\!=\!3$ onward. The two angle
bins $(\theta_c,\theta_f)$ of the \texttt{ANGLE} token (stored in $h_0,h_1$)
parameterize the in-plane rotational freedom that is not pinned by the
three-atom geometry alone (Appendix~\ref{app:frame}). Because the Btriple
is a given prefix excluded from the loss (\S\ref{sec:experiments}), the
model never has to predict these three frames. At generation time, the
Btriple is sampled from a frame cache so that the three positions are
non-collinear, guaranteeing a well-defined $F_{\mathrm{boot}}$. As a minor
consequence, purely one-dimensional (strictly collinear) molecules cannot be
represented by this tokenization: a non-degenerate bootstrap frame requires
three non-collinear atoms.

\paragraph{Placement tokens use the parent's stored frame.}
An \texttt{ADD} step bonds a new atom $A_n$ to its parent $A_{\pi(n)}$. The
bond $p_n - p_{\pi(n)}$ is written in the parent's stored frame $F_{\pi(n)}$
as a distance $r$ and a direction $\hat{u}$:
\begin{equation}
\begin{aligned}
  r &= \|p_n - p_{\pi(n)}\|, \\
  \hat{u} &= F_{\pi(n)}\,\frac{p_n - p_{\pi(n)}}{r} \;\in\; S^2, \\
  p_n &= p_{\pi(n)} + r\,F_{\pi(n)}^{\top}\hat{u}.
\end{aligned}
  \label{eq:localplace}
\end{equation}
The first two identities encode the token during training; the third inverts
them at generation, reusing $F_{\pi(n)}$, built from the parent's ancestor
edges (Eqs.~\ref{eq:uvec}--\ref{eq:gs-frame}) when the parent was placed.

\texttt{LINK} encodes a bond the same way, but between two already-placed atoms
(a ring closure) and adds no new atom. Because both atoms already have
positions, the $(r,\hat{u})$ of a \texttt{LINK} token is redundant;
keeping the redundant $(r,\hat{u})$ is expected to make the learned local-frame
chain more robust.

\paragraph{Collinear and bootstrap fallbacks.}
If $u_1 \parallel u_2$ (e.g.\ along a linear chain), we walk up the ancestor
chain, replacing $u_1\!\leftarrow\!u_2$ and taking the next edge above as the
new $u_2$, until a non-collinear pair is found. If the chain is exhausted we
fall back to the axes of the bootstrap frame (and ultimately to a fixed world
axis) as $u_2$; these exotic cases are { rarely} triggered in practice on the
GEOM-Drugs molecules we use.

\paragraph{SE(3)-invariance.}
A global rigid transformation $p_k\mapsto Rp_k + t$ rotates both $u_1$ and
$u_2$ by $R$, so every stored frame $F_k$ is replaced by $F_k R^{\top}$. The
tokenized pair $(r,\hat{u})$ in Eq.~\eqref{eq:localplace} is therefore
unchanged and the whole token sequence is invariant. SE(3)-invariance thus
arises by construction from the local-frame tokenization, without any
equivariant layers in the transformer. The Btriple (steps 0--2)
establishes the initial frame; the angle bins $(\theta_c,\theta_f)$ of the
\texttt{ANGLE} step parameterize the in-plane rotation that completes the
bootstrap (Appendix~\ref{app:frame}).

\subsection{Discrete token layout}
\label{sec:tokenization}

Given the local-frame placement (Eq.~\ref{eq:localplace}), each non-terminal
atomic event is encoded as a token step made of seven individual tokens:
\[
  t = [a,\,f,\,Z,\,r_b,\,h_0,\,h_1,\,h_2].
\]
Here $a$ is the action --- one of \texttt{INIT}, \texttt{CHAIN},
\texttt{ANGLE}, \texttt{ADD}, \texttt{LINK}, \texttt{END} --- $f$ is the
offset pointer to the parent
(\S\ref{sec:offset}), $Z$ is the new atom's atomic number, $r_b$ discretizes
the bond distance $r$ via a log-spaced binning (Eq.~\ref{eq:rbin},
$R_{\min}{=}0.80\,\angstrom$, $R_{\max}{=}2.50\,\angstrom$, $B{=}200$ bins), and
$(h_0,h_1,h_2)$ encodes the local-frame direction $\hat{u}$ via a 3-slot
factorization of the HEALPix pixel index at $N_{\mathrm{side}}{=}16$
(Eq.~\ref{eq:hpix}, $12{\times}16{\times}16{=}3072$ equal-area pixels). For \texttt{ANGLE} the slots $h_0,h_1$ are reused to hold the two
angle bins $(\theta_c,\theta_f)$ that complete the bootstrap frame; for
\texttt{LINK} the slot $Z$ is reused to encode the offset of the closure
partner (Table~\ref{tab:toksample}).

Let $N_{\mathrm{step}}$ denote the total number of token steps in the sequence,
i.e.\ the Btriple plus every \texttt{ADD}/\texttt{LINK} step plus the
final \texttt{END}. The first $N_{\mathrm{step}}-1$ token steps each contribute
7 tokens; the last step (\texttt{END}) contributes a single token. The total
token count is therefore
\begin{equation}
  N \;=\; 7\,(N_{\mathrm{step}}-1) \,+\, 1.
  \label{eq:Ntokens}
\end{equation}
For the 4-aminophenol example in Table~\ref{tab:toksample},
$N_{\mathrm{step}}=10$ and $N=64$.

\subsection{Offset pointer}
\label{sec:offset}

Parent references use a relative offset, defined as follows. At the step
that places a new atom $A_k$, let $N=k$ be the count of atoms already placed
($A_0,\dots,A_{k-1}$). A parent $A_j$ ($0\leq j\leq k-1$) is addressed by
\[
  f \;=\; -\bigl(N - j\bigr) \;\in\; \{-1, -2, \ldots, -N\},
\]
so $f=-1$ points to the most recently placed atom $A_{k-1}$, $f=-2$ to
$A_{k-2}$, and so on. This is relative rather than absolute: the value
of $f$ does not depend on the global sequence position, only on how many atoms
back the parent lies. The key point is that pointing to an ancestor by ``how
many atoms back'' decouples the pointer distribution from the absolute length
of the sequence: once the model has moved past the bootstrap region, the
statistics of $f$ depend only on local tree topology (typical bond-graph
distance to the parent), not on where the current atom sits in the global
sequence. This gives the model a form of translational invariance along the
sequence axis, the same learnt mapping from local context to $f$ applies
whether we are at step~20 or step~200.

For simplicity, our current implementation truncates $f$ to the range
$\{-1,-2,\ldots,-50\}$ and predicts it with a single $50$-way softmax
($\texttt{max\_offset}=50$). This truncation is a practical convenience, not a
property of the formulation: because $f$ is relative, a factored or
hierarchical softmax (e.g.\ two digit-wise heads) can cover much larger offsets
with a small, fixed parameter budget, and we expect to drop the hard
$\texttt{max\_offset}$ bound in subsequent versions.

\section{Experiments}
\label{sec:experiments}

\paragraph{Architecture and training recipe.}
We use the causal transformer 
(Figure~\ref{fig:pipeline}) with $d=768$ hidden size, $12$ layers, $85.7$M
parameters, standard pre-LayerNorm blocks and multi-head self-attention,
implemented in PyTorch. The encoder is a token-embedding table that maps
each of the slot vocabularies (action, offset, atom type, $r_b$, $h_0$, $h_1$,
$h_2$) to a 768-dim vector; the decoder is an output head that projects
the final transformer hidden state back to a softmax over the appropriate slot
vocabulary. All slot-level losses are cross-entropy, including the distance
bin $r_b$, which is treated as a standard 200-way classification problem (no
continuous/regression loss is used).

\paragraph{No positional encoding.}
\adt{} uses no explicit positional encoding (no sinusoidal,
no learned absolute position, no RoPE). Spatial information is
already supplied at two levels by the tokenization itself:
(i)~the offset slot encodes a relative pointer to the parent
atom in placement order, and the parent's local frame fixes the
direction of the new atom; (ii)~the 7-slot token cycle
(action, offset, $Z$, $r_b$, $h_0$, $h_1$, $h_2$) means that the
position within an atom's token group is determined by the
absolute index modulo 7, which the model can read off from the
slot-vocabulary embeddings without any additional position signal.
Causal self-attention with no positional encoding then suffices;
this matches our preferred reading of \adt{} as ``invariance is in
the tokens, not in the network''.

\paragraph{Random-frontier free-order training.}
A given molecule does not have a single canonical token sequence:
any rooted spanning tree induces a valid tokenization, and there are
combinatorially many of them. We exploit this: \adt{} is trained with
random-frontier free-order augmentation. For every training
sample at every epoch, a fresh tokenization is generated by (i)
drawing the Btriple uniformly at random from the molecule
(which fixes the root and the initial local frame), and (ii)
repeatedly drawing the next atom uniformly at random from the
frontier of unplaced atoms adjacent to the already-placed
set, until every atom has been added (this is neither pure DFS nor
pure BFS but a random-frontier expansion). Different orderings of the same
molecule produce different token sequences but encode the
same 3D structure, so the model is implicitly trained on a
large equivalence class per molecule. Because each training root is a
random heavy atom, the model also sees many scaffolds as initial
contexts; this is what lets a fixed scaffold prefix condition
generation at inference with no scaffold-specific retraining
(\S\ref{sec:drugs30}). The Btriple is
treated as a no-loss prefix, so the cross-entropy loss is computed
only on steps~3 onward (\S\ref{sec:localframe}). We use AdamW.

\paragraph{Token-level AR generation.}
Generation is AR at the token level: within each token step
the model emits the seven constituent tokens one after another, with
temperature $T{=}1.0$, and the model then advances to the next token step. The
prefix fed into the model always contains at least the Btriple (steps
0--2), which fixes the initial local frame. If the prefix is a scaffold with
three or more atoms (e.g.\ benzene, pyridine; \S\ref{sec:drugs30}), the
scaffold already contains the Btriple as its first three tokenization
steps, and any additional scaffold atoms become further given-prefix token
steps; no separate Btriple selection is required. When no specific scaffold is
desired (unconditional generation), a Btriple drawn
at random from a dataset-level frame cache serves as the prefix. The
\texttt{END} token terminates the sequence.

\subsection{Molecule evaluation protocol} \label{sec:xvr}

To evaluate generated heavy-atom-only skeleton molecules $X$ 
given by the token streams of \adt{}, we take two steps:
\begin{enumerate}
  \item[(i)] Completing $X$ with hydrogens into a full molecule, needed because xTB cannot relax a bare heavy-atom skeleton.
  \item[(ii)] Checking whether xTB relaxation preserves $\badt$.
\end{enumerate}
Here $X$ means a set of atomic species, positions and bonds. The set specifies a skeleton molecule.
Step (ii) means
\begin{equation}
  {\bper{X_{\mathrm{relaxed}}} \;=\; \badt} .
  \label{eq:xtp}
\end{equation}
where $G(X)$ is the graph determined from the atomic positions of $X$ (Appendix~\ref{app:topology}).

\begin{figure*}[t!]
  \centering
  \begin{tikzpicture}[
      font=\footnotesize,
      box/.style={draw, rectangle, rounded corners=3pt, align=left,
          inner sep=5pt, minimum height=0.66cm},
      inp/.style={box, fill=gray!12, minimum width=5.8cm},
      stg/.style={box, fill=cyan!12, minimum width=5.8cm},
      xtb/.style={box, fill=green!14, minimum width=5.8cm},
      pass/.style={box, fill=green!32, minimum width=5.8cm, very thick},
      rej/.style={box, fill=red!8, minimum width=5.4cm, align=left, font=\scriptsize},
      arr/.style={-Stealth, thick, color=blue!55!black},
      rarr/.style={-Stealth, color=red!45!black, densely dashed},
      farr/.style={-Stealth, very thick, color=red!78!black},
    ]
    \node[inp] (gen) at (0,0)      {ADT heavy-atom output};
    \node[stg] (nc)  at (0,-1.45)  {\textbf{noclash}\,: $\badt$ connected $+$ no steric clash};
    \node[stg] (ml)  at (0,-2.90)  {\textbf{MLnH}\,: machine-learned $n_{\mathrm{H}}$};
    \node[stg] (mh)  at (0,-4.35)  {\textbf{MLHplacer}\,: machine-learned hydrogen placement};
    \node[xtb] (hp)  at (0,-6.15)  {\textbf{H-prerelax}\,: heavy frozen $\Rightarrow E_{\mathrm{Hprerx}}$};
    \node[xtb] (tp)  at (0,-7.60)  {\textbf{TP Relax}\,: relax keeping $\badt$ $\Rightarrow E_{\mathrm{fullrx}}$};
    \node[align=center] (xt) at (0,-9.35)
      {\textbf{XTP} pass\\[1pt]${\it \Delta} E_{\text{xTB}}= E_{\mathrm{Hprerx}}-E_{\mathrm{fullrx}}$};
    \draw[arr] (gen) -- (nc)
      node[midway, left=3pt, align=right, font=\scriptsize\itshape, text=gray!45!black]
      {the \textbf{ADT topology} $\badt$\\(ADD parent bonds\\$+$ LINK ring closures)};
    \draw[arr] (nc) -- (ml);  \draw[arr] (ml) -- (mh);  \draw[arr] (mh) -- (hp);
    \draw[arr] (hp) -- (tp);
    \draw[arr] (tp) -- (xt)
      node[midway, right=3pt, font=\footnotesize] {$N_{\mathrm{XTP}}$};

    \node[anchor=west, font=\footnotesize] at (0.4,-0.5)  {$N$};
    \node[anchor=west, font=\footnotesize] at (0.4,-1.95) {$N_{\mathrm{noclash}}$};
    \node[anchor=west, font=\footnotesize] at (0.4,-3.40) {$N_{n\mathrm{H}}$};
    \node[anchor=west, font=\footnotesize] at (0.4,-4.85) {$N_{\mathrm{Hplacer}}$};
    \node[anchor=west, font=\footnotesize] at (0.4,-6.65) {$N_{\mathrm{Hprerx}}$};
    \node[anchor=west, font=\footnotesize] at (0.4,-8.10) {$N_{\mathrm{fullrx}}$};

    \draw[farr, rounded corners=3pt] (tp.west) -- (-4.2,-7.60) -- (-4.2,-5.22) -- (0,-5.22);
    \draw[farr, rounded corners=3pt] (hp.west) -- (-3.5,-6.0) -- (-3.5,-5.44) -- (0,-5.44);
    \node[align=left, font=\scriptsize, text=red!72!black] at (-5.9,-6.9)
       {\textbf{H-integrity fallback}:\\Strip even-detached H};
  \end{tikzpicture}
  \caption{\textbf{The molecule-evaluation funnel} (protocol \S\ref{sec:xvr}).
    A batch of $N$ generated heavy-atom skeletons descends a hydrogen-completion phase
    (\textbf{noclash}, \textbf{MLnH}, \textbf{MLHplacer}) and a relaxation-and-acceptance
    phase (\textbf{H-prerelax}, \textbf{TP Relax}) to the \textbf{XTP}-accepted set
    $N_{\mathrm{XTP}}$; the number beside each arrow is that stage's survivor count, and the
    two learned stages are lossless
    ($N_{\mathrm{noclash}}{=}N_{n\mathrm{H}}{=}N_{\mathrm{Hplacer}}$).
    Acceptance is tested against the ADT topology $\badt$ the model emitted
    (\textsc{add} parent bonds $+$ \textsc{link} ring closures), read from the 3D coordinates
    alone with no SMILES. Left (red): the \textbf{H-integrity fallback} restarts an even
    hydrogen detachment and rejects an odd one.}
  \label{fig:xvr-flow}
\end{figure*}
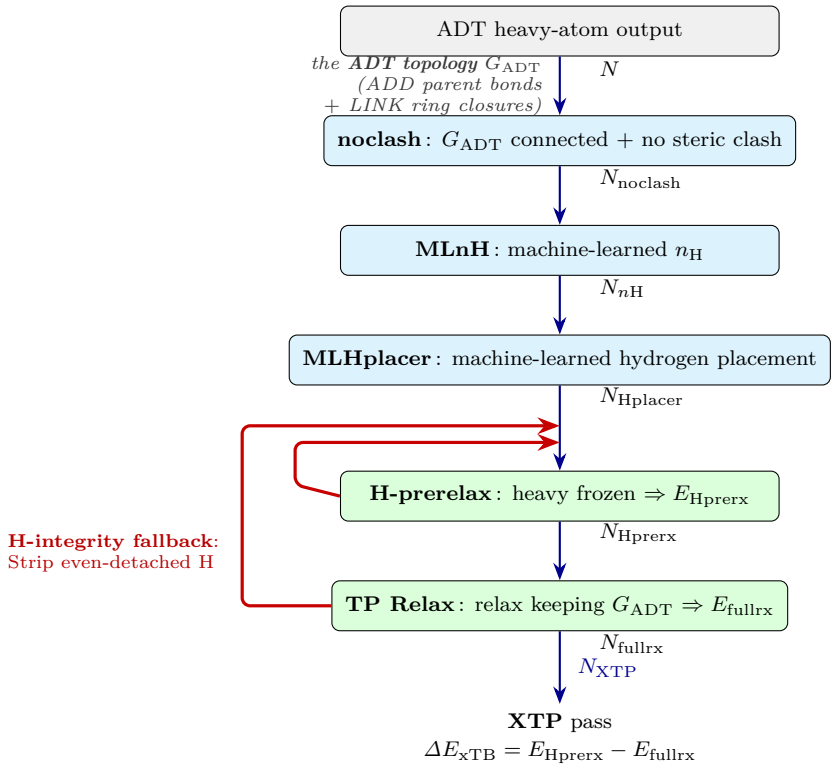

Figure~\ref{fig:xvr-flow} shows the full funnel; the two phases are detailed next.
\paragraph{Phase 1 --- hydrogen completion (heavy-atom-only skeleton molecule $\to$ full molecule).}
A generated sample is a bare heavy-atom-only skeleton molecule, which GFN2-xTB cannot relax; three steps complete it
with hydrogens into a full 3D molecule ready for relaxation. \textbf{noclash} keeps
the $N_{\mathrm{noclash}}$ samples whose heavy atoms form a single connected component with no steric
clash. \textbf{MLnH} then assigns each heavy atom a hydrogen count with a machine-learned completer,
followed by a $\pm1$\,H parity correction that enforces an even electron count, consistent with a neutral closed-shell xTB calculation.
Then \textbf{MLHplacer} positions those hydrogens in 3D.
Both stages are robust enough to keep $N_{\mathrm{noclash}}{=}N_{n\mathrm{H}}{=}N_{\mathrm{Hplacer}}$.
Thus the only attrition in this phase is the geometric \textbf{noclash} filter.
Both MLnH and MLHplacer are transformers. See Appendix~\ref{app:hcomplete}.

\paragraph{Phase 2 --- relaxation and acceptance.}
The hydrogenated molecule is relaxed with GFN2-xTB in two steps. \textbf{H-prerelax} relaxes only the
hydrogens, the heavy atoms frozen, giving a hydrogen-placement energy $E_{\mathrm{Hprerx}}$
($N_{\mathrm{Hprerx}}$ converge). \textbf{TP Relax (topology-preserving Relax)} then relaxes the heavy atoms in two stages: first
with the declared $\badt$ bond distances restrained, so that $\badt$ is realized; then with the
restraints released, relaxing freely to $E_{\mathrm{fullrx}}$ ($N_{\mathrm{fullrx}}$ converge). The free
stage tests whether $\badt$ sits at a genuine minimum, or was held only by the restraints. A sample is
XTP-accepted iff the freely relaxed geometry satisfies Eq.~\eqref{eq:xtp}; the accepted count is
$N_{\mathrm{XTP}}$.

A hydrogen can drift off during relaxation. An even detachment is stripped, the molecule restarting from
H-prerelax; an odd detachment leaves a radical, which is rejected (the H-integrity fallback;
Figure~\ref{fig:xvr-flow}, left). This fallback is needed to rescue a few percent of generated molecules.\\

The relaxation energy $\Delta E_{\mathrm{xTB}}{=}E_{\mathrm{Hprerx}}{-}E_{\mathrm{fullrx}}$ is a per-molecule
strain, measuring how far the generated geometry sat from the xTB minimum. The survivor counts
($N_{\mathrm{noclash}}\!\ge\!N_{\mathrm{Hprerx}}\!\ge\!N_{\mathrm{fullrx}}\!\ge\!N_{\mathrm{XTP}}$) and
the strain are tabulated in Section~\ref{sec:drugs30}.

\subsection{Pretrained model} \label{sec:drugs30}

We center the evaluation on GEOM-Drugs, whose GFN2-xTB-optimized conformers are
methodologically consistent with our physical (strain\,$+$\,\xvr{}) axis and are drug-relevant.

\begin{figure}[t!]
  \centering
  \begin{tikzpicture}
    \begin{axis}[
        width=8.4cm, height=3.4cm,
        ybar,
        bar width=1.3mm,
        bar shift=0pt,
        xlabel={\footnotesize heavy-atom count},
        ylabel={\footnotesize \# molecules ($\times 10^3$)},
        xmin=0, xmax=51.5, ymin=0, ymax=23,
        xtick={0,5,10,15,20,25,30,35,40,45,50},
        ytick={0,5,10,15,20},
        tick label style={font=\scriptsize},
        label style={font=\scriptsize},
        enlarge x limits=false,
        axis on top,
      ]
      \addplot[fill=blue!20, draw=blue!55!black] coordinates {
          (3,0.004) (4,0.006) (5,0.024) (6,0.040) (7,0.081)
          (8,0.118) (9,0.168) (10,0.388) (11,0.644) (12,1.052)
          (13,1.655) (14,2.740) (15,4.313) (16,6.070) (17,8.509)
          (18,11.004) (19,13.758) (20,16.489) (21,18.828) (22,20.209)
          (23,21.185) (24,21.740) (25,21.401) (26,20.664) (27,19.561)
          (28,18.072) (29,15.920) (30,13.838) (31,11.404) (32,9.276)
          (33,7.434) (34,5.373) (35,3.932) (36,2.589) (37,1.600)
          (38,1.133) (39,0.733) (40,0.568) (41,0.377) (42,0.314)
          (43,0.167) (44,0.134) (45,0.070) (46,0.085) (47,0.054)
          (48,0.033) (49,0.033) (50,0.030)
        };
    \end{axis}
  \end{tikzpicture}
  \caption{GEOM-Drugs heavy-atom distribution of the full $304{,}335$-molecule reference set
    (one conformer per SMILES, GFN2-xTB-optimized as released in GEOM), mean $24.86\pm 5.67$;
    $84.9\%$ of molecules have $\le\!30$ heavy atoms (the pretraining cut).}
  \label{fig:drugs-dist}
\end{figure}
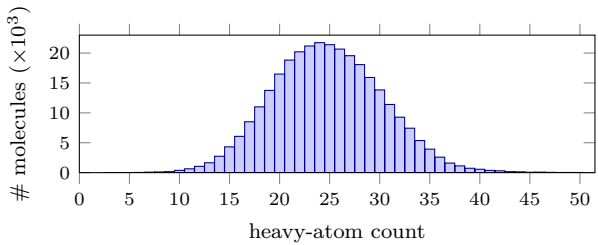

\paragraph{Data.}
Our reference dataset is the drug-like ``Drugs'' subset of GEOM~\citep{geom2022}, downloaded as
the \texttt{rdkit\_folder} release from the Harvard Dataverse~\citep{geomdataverse}.
Taking one conformer (the first one listed in the release, a GFN2-xTB-optimized conformer that is the lowest-energy one for about $82\%$ of molecules) per unique SMILES, 
the release yields $304{,}335$ valid RDKit molecules. We call this the GEOM-Drugs database throughout.
See Figure~\ref{fig:drugs-dist}.
For the pretraining, we restrict to heavy-atom count $\leq 30$, retaining $257{,}574$ molecules ($84.6\%$
  of the full set, mean $23.28$); this cut keeps the model focused on the bulk of
drug-like chemistry while excluding the long tail of larger molecules. 

\paragraph{Training.}
We train from scratch with 2$\times$RTX~4090 at $\sim 365$\,s per epoch; 
the checkpoint used for generation is E240 (after 240 epochs
of training, $\sim 24$\,h wall clock).

\paragraph{Generation.}
Generation starts the
autoregressive decoder from a bootstrap frame sampled from a frame cache
(a pool of initial tokenized steps extracted from GEOM training molecules)
after which the model completes the molecule one atom at a time. We use two
modes. One is the unconditional generation
(the Btriple (uncond.) row of Table~\ref{tab:drugs30}), where  we draw a random
valid Btriple, which is the first three tokenized steps of a training
molecule. Then the model is free to grow any topology.
The other is the scaffold-conditional generation, where a per-scaffold cache supplies
bootstrap frames whose opening steps lay down a chosen ring taken directly from
real GEOM structures (benzene, pyridine, pyrimidine, pyrazine, furan, thiophene,
or cyclohexane in our experiments); fixing such a frame as the prefix forces
every completion to contain that scaffold. A single trained model handles all
scaffolds without scaffold-specific retraining: the scaffold choice is purely
an inference-time conditioning decision. 
Unconditional generation is thus scaffold-conditional generation whose prefix is only a random Btriple.

\paragraph{Results.}
The \textbf{Pretrained} block of Table~\ref{tab:drugs30} reports scaffold-conditional
generation with the seven ring prefixes, $N{=}10{,}000$ each. Averaged over scaffolds,
$N_{\mathrm{Hprerx}}{=}7193$ molecules converge in the H-prerelax; a quarter then change topology under
the free relaxation, leaving $N_{\mathrm{XTP}}{=}5547$ ($55.5\%$) XTP-accepted. A further RDKit check
counts those that also read back as a valid neutral molecule:
$N_{\mathrm{XTP}}^{\mathrm{smiles}}{=}5355$, $96.5\%$ of the accepted set.
After removing duplicates, the end-to-end yield of distinct valid molecules
is $N^{\mathrm{gen}}/N{=}52.6\%$ (unconditional Btriple $41.8\%$). The XTP-accepted geometries sit near,
not tight against, the GFN2-xTB minimum: median Kabsch RMSD $0.47$\,\AA{} and median strain
$28.9$\,kcal\,mol$^{-1}$. Generation on one RTX~4090 requires $\sim\!20$\,ms per atom
($\sim\!0.6$\,s for a 30-atom molecule), sample-by-sample without batching.

The XTP-accepted molecules are almost all distinct and new: novelty (canonical SMILES absent from the
full GEOM-Drugs set, hence also from the training set) is $\geq$ $99.1\%$ for every scaffold, so the model neither
collapses onto a few modes nor memorizes the data. Chemical diversity matches GEOM-Drugs sub-sampled to
the same count (Table~\ref{tab:divers30}): the number of distinct Bemis--Murcko
scaffolds~\citep{bemismurcko1996} meets or exceeds the reference, and internal diversity
IntDiv$_1$~\citep{polykovskiy2020moses} is $\sim\!0.85$--$0.89$. Molecular weight, $\log P$ and QED sit
in drug-like ranges (Table~\ref{tab:divers30}).
\begin{table*}[t!]
  \centering
  \small
  \setlength{\tabcolsep}{3pt}
  \begin{tabular}{l rrrr r @{\hskip 10pt} r @{\hskip 8pt} c rr}
    \toprule
    & \multicolumn{4}{c}{validity funnel (survivors of $N{=}10^4$)} & & & & \multicolumn{2}{c}{geometry} \\
    \cmidrule(lr){2-5}\cmidrule(lr){9-10}
    \textbf{Scaffold} & $N_{\mathrm{Hprerx}}$ & $N_{\mathrm{XTP}}$ & $N_{\mathrm{XTP}}^{\mathrm{smiles}}$ & $N^{\mathrm{gen}}$\,(rate) & $N^{\mathrm{Novel}}$ & $N_{\mathrm{fullrx}}^{\dagger}$ & \shortstack{\scriptsize size (heavy)\\ \scriptsize mean/std/med/min/max} & \shortstack{RMSD\\(\AA)} & \shortstack{$\Delta E$\\ \scriptsize kcal/mol} \\
    \midrule
    \multicolumn{10}{@{}l}{\textbf{Pretrained model}} \\[1pt]
    benzene & 7131 & 5535 & 5313 & 5291\,(52.9\%) & 5244 & 7028 & \scriptsize 26.6/6.3/28/7/35 & 0.49 & 30.4 \\
    pyrimidine & 6990 & 5400 & 5113 & 5066\,(50.7\%) & 5045 & 6900 & \scriptsize 27.1/6.1/29/7/35 & 0.45 & 27.4 \\
    furan & 7167 & 5473 & 5302 & 5252\,(52.5\%) & 5232 & 7089 & \scriptsize 26.4/6.4/28/7/35 & 0.51 & 31.8 \\
    pyridine & 7394 & 5840 & 5635 & 5583\,(55.8\%) & 5542 & 7318 & \scriptsize 25.7/6.6/27/7/36 & 0.42 & 25.1 \\
    thiophene & 7047 & 5290 & 5208 & 5176\,(51.8\%) & 5154 & 6937 & \scriptsize 25.9/6.4/28/7/35 & 0.47 & 30.8 \\
    pyrazine & 7641 & 6224 & 5973 & 5574\,(55.7\%) & 5551 & 7570 & \scriptsize 24.7/7.3/26/7/35 & 0.43 & 24.1 \\
    cyclohexane & 6984 & 5068 & 4944 & 4880\,(48.8\%) & 4850 & 6831 & \scriptsize 26.3/6.4/28/6/35 & 0.51 & 32.6 \\
    \addlinespace[1pt]
    \textbf{average} & \textbf{7193} & \textbf{5547} & \textbf{5355} & \textbf{5260\,(52.6\%)} & \textbf{5231} & \textbf{7096} & \scriptsize \textbf{26.1/6.5/28/7/35} & \textbf{0.47} & \textbf{28.9} \\
    Btriple (uncond.) & 6307 & 4454 & 4182 & 4182\,(41.8\%) & 4172 & 6168 & \scriptsize 27.0/6.4/29/4/35 & 0.53 & 34.8 \\
    \midrule
    \multicolumn{10}{@{}l}{\textbf{RLVR model}} \\[1pt]
    benzene & 9961 & 9834 & 9636 & 9588\,(95.9\%) & 9544 & 9958 & {\scriptsize 25.9/5.4/26/7/46} & 0.24 & 9.8 \\
    pyrimidine & 9953 & 9830 & 9650 & 9574\,(95.7\%) & 9560 & 9943 & {\scriptsize 25.0/5.4/25/9/48} & 0.23 & 9.5 \\
    furan & 9957 & 9859 & 9746 & 9565\,(95.7\%) & 9520 & 9954 & {\scriptsize 24.8/5.4/25/7/47} & 0.24 & 8.7 \\
    pyridine & 9968 & 9857 & 9712 & 9608\,(96.1\%) & 9572 & 9968 & {\scriptsize 25.3/5.5/25/9/44} & 0.23 & 8.5 \\
    thiophene & 9975 & 9866 & 9835 & 9636\,(96.4\%) & 9619 & 9972 & {\scriptsize 24.7/5.5/25/8/45} & 0.23 & 8.6 \\
    pyrazine & 9963 & 9857 & 9489 & 9131\,(91.3\%) & 9122 & 9959 & {\scriptsize 24.9/5.6/25/6/47} & 0.23 & 9.5 \\
    cyclohexane & 9969 & 9844 & 9788 & 9242\,(92.4\%) & 9208 & 9967 & {\scriptsize 25.3/5.5/25/10/46} & 0.21 & 7.6 \\
    \addlinespace[1pt]
    \textbf{average} & \textbf{9964} & \textbf{9850} & \textbf{9694} & \textbf{9478\,(94.8\%)} & \textbf{9449} & \textbf{9960} & {\scriptsize \textbf{25.1/5.5/25/8/46}} & \textbf{0.23} & \textbf{8.9} \\
    Btriple (uncond.) & 9951 & 9819 & 9675 & 9673\,(96.7\%) & 9635 & 9946 & {\scriptsize 25.4/5.5/25/6/45} & 0.26 & 8.6 \\
    \bottomrule
  \end{tabular}
  \caption{\textbf{The molecule-evaluation funnel, in numbers.}
  Survivor counts along the funnel of Figure~\ref{fig:xvr-flow}, for the pretrained model (upper block)
  and the RLVR model (lower block). $N{=}10{,}000$ per scaffold.
  $N_{\mathrm{Hprerx}}$: molecules that relax with the attached hydrogens (heavy atoms frozen).
  $N_{\mathrm{XTP}}$: the accepted set, topology-preserving after relaxation.
  $N_{\mathrm{XTP}}^{\mathrm{smiles}}$: those RDKit reads back as the declared molecule (a neutral Lewis structure whose heavy-atom bonds are exactly the declared ones).
  $N^{\mathrm{gen}}$: distinct molecules among $N_{\mathrm{XTP}}^{\mathrm{smiles}}$.
  $N^{\mathrm{Novel}}$: those absent from the full GEOM-Drugs set.
  $N_{\mathrm{fullrx}}^{\dagger}$: molecules whose free relaxation converged (a diagnostic count, off the funnel).
  RMSD and $\Delta E_{\mathrm{xTB}}{=}E_{\mathrm{Hprerx}}{-}E_{\mathrm{fullrx}}$ are the median heavy-atom
  Kabsch RMSD and relaxation strain over the XTP-accepted set. \textbf{size} is its heavy-atom-count
  distribution. The average row is the per-scaffold mean. The Btriple (uncond.) row is unconditional
  generation, excluded from that average.}
  \label{tab:drugs30}
\end{table*}
\begin{table*}[t!]
\centering\small
\setlength{\tabcolsep}{3.5pt}
\begin{tabular}{lrcccrrrr}
\toprule
 & & \multicolumn{3}{c}{diversity\quad(model\,/\,GEOM at the same $N^{\mathrm{gen}}$)} & \multicolumn{4}{c}{properties (model)} \\
\cmidrule(lr){3-5}\cmidrule(lr){6-9}
\textbf{Scaffold} & $N^{\mathrm{gen}}$ & $N_{\mathrm{eff}}^{\mathrm{scaf}}$ & \#distinct & IntDiv$_1$ & MW (Da) & $\log P$ & QED & \shortstack{$\Delta E_{\text{xTB}}$\\(kcal\,mol$^{-1}$)} \\
\midrule
\multicolumn{9}{@{}l}{\textbf{Pretrained model}} \\[1pt]
benzene & 5291 & 3531/3593 & 4648/4376 & 0.891/0.868 & $384.6\pm90.4$ & $+2.99\pm1.57$ & $0.505\pm0.199$ & $30.4$ \\
pyridine & 5583 & 3364/2901 & 4785/3887 & 0.881/0.862 & $370.6\pm93.7$ & $+2.54\pm1.61$ & $0.507\pm0.199$ & $25.1$ \\
pyrimidine & 5066 & 3064/2569 & 4204/3353 & 0.892/0.864 & $393.0\pm86.2$ & $+2.26\pm1.80$ & $0.460\pm0.192$ & $27.4$ \\
pyrazine & 5574 & 2524/826 & 4298/1524 & 0.847/0.850 & $364.8\pm91.7$ & $+1.88\pm1.57$ & $0.522\pm0.197$ & $24.1$ \\
furan & 5252 & 3751/2630 & 4663/3432 & 0.860/0.836 & $383.4\pm90.2$ & $+2.68\pm1.63$ & $0.491\pm0.201$ & $31.8$ \\
thiophene & 5176 & 3860/2589 & 4608/3451 & 0.887/0.851 & $391.0\pm91.1$ & $+2.23\pm1.91$ & $0.414\pm0.196$ & $30.8$ \\
cyclohexane & 4880 & 3248/2366 & 4259/3199 & 0.860/0.853 & $380.8\pm89.9$ & $+2.90\pm1.56$ & $0.548\pm0.195$ & $32.6$ \\
\addlinespace[2pt]
Btriple (uncond.) & 4182 & 3030/3090 & 3735/3638 & 0.897/0.873 & $393.6\pm92.8$ & $+2.40\pm1.80$ & $0.473\pm0.200$ & $34.8$ \\
\midrule
\multicolumn{9}{@{}l}{\textbf{RLVR model}} \\[1pt]
benzene & 9588 & 3890/5734 & 6247/7427 & 0.845/0.868 & $367.9\pm79.6$ & $+4.14\pm1.69$ & $0.495\pm0.216$ & 9.8 \\
pyridine & 9608 & 3814/4079 & 6312/5928 & 0.835/0.862 & $358.7\pm80.2$ & $+3.78\pm1.74$ & $0.514\pm0.213$ & 8.5 \\
pyrimidine & 9574 & 2395/3611 & 5180/5329 & 0.854/0.864 & $357.1\pm79.8$ & $+3.24\pm1.86$ & $0.504\pm0.201$ & 9.5 \\
pyrazine & 9131 & 1688/826 & 4483/1524 & 0.815/0.850 & $359.4\pm80.1$ & $+3.60\pm1.74$ & $0.527\pm0.210$ & 9.5 \\
furan & 9565 & 3035/3682 & 5685/5304 & 0.820/0.836 & $354.9\pm78.8$ & $+3.82\pm1.71$ & $0.527\pm0.210$ & 8.7 \\
thiophene & 9636 & 2902/3697 & 5695/5415 & 0.846/0.851 & $368.9\pm79.1$ & $+3.32\pm2.13$ & $0.425\pm0.217$ & 8.6 \\
cyclohexane & 9242 & 2228/3342 & 4849/5138 & 0.781/0.853 & $364.5\pm76.9$ & $+4.19\pm1.69$ & $0.588\pm0.226$ & 7.6 \\
\addlinespace[2pt]
Btriple (uncond.) & 9673 & 5461/5937 & 7351/7646 & 0.856/0.873 & $364.0\pm80.5$ & $+3.61\pm1.83$ & $0.521\pm0.220$ & 8.6 \\
\bottomrule
\end{tabular}
\caption{\textbf{Per-scaffold diversity and properties, compared to GEOM-Drugs.} 
Pretrained (upper block) vs.\ RLVR (lower block). Scaffold-diversity
  measures grow with the number of molecules sampled, so each diversity cell is
  model\,/\,GEOM, where GEOM-Drugs is \emph{randomly sub-sampled to the same molecule count
  $N^{\mathrm{gen}}$} as that row (column~$N^{\mathrm{gen}}$; benzene rows use benzene-containing GEOM, etc.; Btriple vs.\ all
  of GEOM; GEOM holds only $4{,}165$ distinct pyrazine-containing molecules, so the pyrazine GEOM
  values use that whole pool). 
  We report two complementary scaffold measures. \#distinct is the raw count of distinct
  Bemis--Murcko scaffolds~\citep{bemismurcko1996}, structural breadth (how many different
  ring$+$linker frameworks occur). $N_{\mathrm{eff}}^{\mathrm{scaf}}{=}\exp(H)$ is the effective
  number, the exponential of the Shannon entropy $H$ of the scaffold-frequency distribution; it
  down-weights rarely used scaffolds and so measures evenness. Always
  $N_{\mathrm{eff}}^{\mathrm{scaf}}\le\#\text{distinct}$, with equality only for a flat distribution.
  IntDiv$_1$~\citep{polykovskiy2020moses} is the internal diversity $1-\overline{T}$ (one minus
  the mean pairwise Tanimoto similarity~\citep{tanimoto2015} on Morgan
  fingerprints~\citep{morgan2010}, radius 2, 2048 bits; GEOM value is sample-size-robust). MW, Crippen
  $\log P$~\citep{crippen1999} and QED are the model's means $\pm$ std (GEOM reference $\approx$
  MW~$356$, $\log P\,{+}2.9$, QED~$0.65$); $\Delta E_{\mathrm{xTB}}$ is the median relaxation strain per molecule.}
\label{tab:divers30}
\end{table*}

\subsection{RLVR model}
\label{sec:drugs50}

\paragraph{Method.}
Starting from the pretrained model, we perform data-free RLVR training of \adt{} against a
verifiable reward, with no external molecules in the loop. Here data-free means that no molecules serve as training
targets: the model learns only from its own samples, scored by xTB. The bootstrap prefixes are drawn from the
GEOM-Drugs frame cache, and the size and composition control terms below steer toward GEOM-Drugs statistics. The loss has two parts,
\begin{equation}
  \mathcal{L} \;=\; \mathcal{L}_{\mathrm{reward}} \;+\; \mathcal{L}_{\mathrm{ctrl}} .
  \label{eq:rlvr-twopart}
\end{equation}
$\mathcal{L}_{\mathrm{reward}}$ raises the verifiable reward $r_{\tau}$ 
on each generated molecule $\tau$, together with a batch-level scaffold-diversity
floor. Neither is differentiable, so both are trained by RL: the reward re-weights the log-likelihood
of the sampled molecule.
$\mathcal{L}_{\mathrm{ctrl}}$ is a sum of \emph{differentiable} control terms, trained by ordinary
gradient descent, that hold \adt{} on the drug-like manifold: a squared-KL trust region to a reference
model, and distribution targets on the heavy-atom-count moment and the element composition. The trust
region is periodically re-anchored (a ratchet) to escape plateaus.
The reward, the full objective (Eq.~\ref{eq:rlvr-loss}) and the hyperparameters are in
Appendices~\ref{app:rlvr} and~\ref{app:training}.

The other side of this training is that \adt{} supplies its own data. At each step the model generates
a batch of molecules, the xTB verifier scores them, and those scores drive the update; nothing is read
from a dataset. The model improves by proposing structures and keeping what physics accepts, in the
spirit of AlphaZero's self-play with the rules of the game replaced by GFN2-xTB.

\paragraph{Training.}
The RLVR run starts from the pretrained model and improves it against the xTB reward, a batch of
$48$ molecules per step. The step is dominated by xTB rather than by the network (three GFN2
relaxations per molecule). Figure~\ref{fig:training-curves}(b) shows the topology-preservation rate
climbing over the run; hyperparameters and compute are in Appendix~\ref{app:training}.

\paragraph{Results.}

\begin{table*}[t!]
  \centering
  \small
  \setlength{\tabcolsep}{6pt}
  \begin{tabular}{lcc}
    \toprule
    Metric & RLVR model & GEOM-Drugs \\
    \midrule
    $N^{\mathrm{gen}}/N$ (distinct valid molecules per sample) & \textbf{96.7\%} & --- \\
    \multicolumn{3}{l}{\textit{Geometric consistency with GFN2-xTB} (GFN2-xTB relaxation strain; reference $0=$ xTB minimum)}\\
    \quad strain / heavy atom (p25 / p50 / p75 / p90, kcal\,mol$^{-1}$) & 0.25 / \textbf{0.35} / 0.50 / 0.82 & $\approx0$ \\
    \quad heavy-atom RMSD (Kabsch), median (\AA) & 0.26 & --- \\
    \quad bond / angle shift, median $|\Delta r|$ / $|\Delta\theta|$ & 0.006\,\AA\ / $1.5^{\circ}$ & --- \\
    \multicolumn{3}{l}{\textit{Diversity \& composition}}\\
    \quad novel (vs.\ full GEOM-Drugs set) & \textbf{99.6\%} & --- \\
    \quad size (heavy), mean\,$\pm$\,std & 25.4\,$\pm$\,5.5 & 24.9\,$\pm$\,5.7 \\
    \quad C / N / O (\%) & 72.49 / 11.20 / 12.30 & 72.61 / 11.64 / 11.74 \\
    \quad S / F / Cl / Br (\%) & 2.07 / 1.03 / 0.71 / 0.18 & 2.18 / 0.88 / 0.73 / 0.19 \\
    \bottomrule
  \end{tabular}
  \caption{\textbf{Summary of the RLVR \adt{} model}, unconditional
  (Btriple) generation: a random valid Btriple, $N{=}10{,}000$ samples for RLVR model; 
  the reference column is the full GEOM-Drugs set
  ($304{,}335$ molecules). $N^{\mathrm{gen}}/N$ is the end-to-end yield of distinct valid molecules. 
  Note that GEOM-Drugs are relaxed by GFN2-xTB.
  RMSD and bond/angle shifts are over the XTP-accepted set, as are strain, size and composition.
  RMSD is Kabsch-aligned and $|\Delta r|$,$|\Delta\theta|$ are the median per-bond and per-angle displacements
  between the generated and relaxed geometry (Figure~\ref{fig:bond-angle-errors}). novel is the
  fraction absent from the full GEOM-Drugs set.}
  \label{tab:drugs30honest}
\end{table*}

In Table~\ref{tab:drugs30honest}, we show a summary of the RLVR model for Btriple (unconditional generation).
Size and composition are well fit to the reference of GEOM-Drugs thanks to $\mathcal{L}_{\mathrm{ctrl}}$.
Such agreement is not automatic. G-SchNet~\citep{gschnet2019}, the closest autoregressive 3D generator, has
been shown to drift away from its training distribution: it over-expresses unsaturated carbon and heteroatoms,
suppresses aliphatic motifs, and does not reproduce the training size distribution~\citep{structuralbias2025}.
The size and composition terms of $\mathcal{L}_{\mathrm{ctrl}}$ are what hold both here.
The size range is emergent, not imposed: the architecture admits up to $56$ heavy atoms and neither
model reaches it (the RLVR model tops out at $45$ in this unconditional run and at $48$ with a scaffold prefix, Table~\ref{tab:drugs30}).

See the \textbf{RLVR} block of Table~\ref{tab:drugs30}.
With no external data, RLVR improves every tabulated stage of the funnel. 
The topology-preserved set $N_{\mathrm{XTP}}$ nearly doubles ($5547\!\to\!9850$; $55.5\%\!\to\!98.5\%$). The end-to-end yield of distinct valid molecules
$N^{\mathrm{gen}}/N$ rises from $52.6\%$ to $\mathbf{94.8\%}$ (unconditional Btriple $41.8\%\!\to\!96.7\%$).
RLVR cleans up the geometry: median Kabsch RMSD falls from $0.47$ to $0.23$\,\AA{} and median relaxation strain from $28.9$ to $8.9$\,kcal\,mol$^{-1}$, so XTP-accepted
molecules already sit close to the GFN2-xTB minimum (Table~\ref{tab:drugs30honest}).
The refined strain is $\sim\!$$0.36$\,kcal\,mol$^{-1}\!\sim\!16$\,meV per atom.
The share of XTP-accepted molecules that RDKit
reads back as the declared molecule ($N_{\mathrm{XTP}}^{\mathrm{smiles}}/N_{\mathrm{XTP}}$) rises from
$96.5\%$ to $98.4\%$; this step tests RDKit-readability, not the existence of a neutral Lewis structure.

Novelty stays $\geq$ $99.5\%$ for every scaffold (Table~\ref{tab:drugs30}), so the refined
model does not memorize. Measured against GEOM-Drugs at a matched sample size
(Table~\ref{tab:divers30}), structural breadth is largely preserved: the number of distinct
Bemis--Murcko scaffolds exceeds the matched-$N$ GEOM reference for four of the seven ring scaffolds (pyridine, pyrazine, furan, thiophene; e.g.\ pyridine $6312$ vs.\ $5928$) and falls short for benzene ($6247$ vs.\ $7427$), pyrimidine ($5180$ vs.\ $5329$), cyclohexane ($4849$ vs.\ $5138$) and the unconditional run ($7351$ vs.\ $7646$), by $3$--$16\%$, while evenness
shows a mild gap: $N_{\mathrm{eff}}^{\mathrm{scaf}}$ is $\sim\!0.66$--$0.94\times$ GEOM (pyrazine, whose GEOM pool is small, lies above it), i.e.\ the
validity reward leans a little harder on a few common ring--linker motifs (diphenyl
ether/ester/amide). Pyridine shows the split cleanly: RLVR has more distinct scaffolds
($6312{>}5928$) but a lower effective count ($3814{<}4079$), using them less evenly.
This evenness gap is a tunable trade-off, not a fixed cost: the scaffold-diversity floor
(Table~\ref{tab:controls}) can be raised, buying evenness at a small validity cost.
Internal diversity IntDiv$_1$ stays high ($\sim\!0.78$--$0.85$, modestly below the
more-scattered pretrained model), and the properties drift mildly toward lipophilicity
(MW $\sim\!360$\,Da, $\log P\sim{+}3.7$, QED $\sim\!0.51$), an uncontrolled, disclosed shift,
not a collapse. 

Because the reward is purely physical (xTB), it does not encode 
the drug-likeness or curation bias of the human-selected GEOM-Drugs set; the RLVR distribution therefore overlaps GEOM in breadth but diverges in curation-driven aspects, 
which we quantify rather than claim GEOM-identity.

Representative generated molecules are listed in Table~\ref{tab:gen-rlvr}. 
Among the $N^{\rm gen}$ generated molecules, $11.0\%$ contain an S$^+$ identified by RDKit, against $3.6\%$ in GEOM-Drugs (the same perception applied to the lowest-energy GFN2-xTB conformers of 5,000 random single-fragment GEOM-Drugs molecules).  
This difference is not a problem, because we do not control the detailed composition of molecules.
A problem is that RDKit assigns $4.0\%$ of the $N^{\rm gen}$ molecules a nonzero formal charge on C, N or O without an S$^+$. A quarter of these ($1.0\%$) are nitro or N-oxide N$^+$/O$^-$ pairs, which RDKit assigns to $4.4\%$ of GEOM-Drugs as well; the remaining $3.0\%$ are ylide-like pairs such as C$^-$/O$^+$, C$^-$/N$^+$ and O$^+$/O$^-$ (block~(c) of Table~\ref{tab:gen-rlvr}), which GEOM-Drugs essentially lacks ($0.08\%$). 
We suspect these are strange structures that xTB admits but that the hand curation of GEOM-Drugs kept
out: xTB, an approximate method with documented limitations~\cite{xtbreview2021}, relaxes them to a
closed-shell minimum realizing the declared graph, and RDKit reads that geometry back as a valid
molecule, so the acceptance rests with the verifier (block~(c) of Table~\ref{tab:gen-rlvr} shows
examples, alongside strained rings such as an in-ring cumulene or an aryne).
Our reward sees only the local relaxation strain and does not explicitly pick out the ground state, so it
does not penalize a molecule that sits at a shallow local minimum yet is high in absolute energy. A total
binding energy term in the reward (for example a composition-normalized atomization energy) would favor
lower-energy structures and suppress this tail, which we leave to future work.

\subsection{Structural fidelity: bond-length and bond-angle errors}
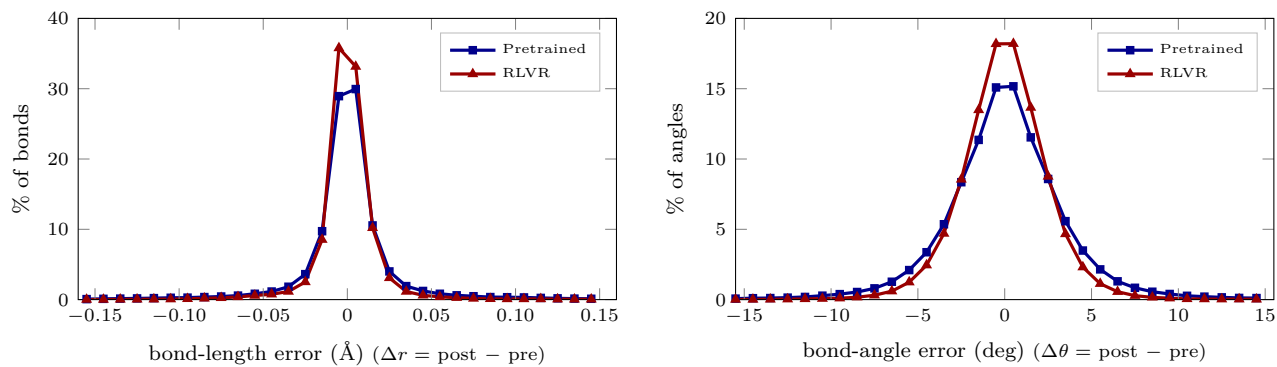
\begin{figure*}[t!]
\centering
\begin{tikzpicture}
\begin{axis}[
    name=plotA,
    width=8.7cm, height=5.3cm,
    xlabel={\footnotesize bond-length error (\AA{}) \scriptsize ($\Delta r =$ post $-$ pre)},
    ylabel={\footnotesize \% of bonds},
    xmin=-0.16, xmax=0.16, ymin=0, ymax=40,
    xtick={-0.15,-0.1,-0.05,0,0.05,0.1,0.15},
    xticklabels={$-0.15$,$-0.10$,$-0.05$,$0$,$0.05$,$0.10$,$0.15$},
    ytick={0,10,20,30,40},
    /pgf/number format/.cd, fixed, precision=2,
    tick label style={font=\scriptsize},
    label style={font=\scriptsize},
    enlarge x limits=false,
    axis on top,
    legend style={font=\tiny, at={(0.97,0.95)}, anchor=north east, draw=gray!50},
    legend cell align=left,
]
\addplot[blue!55!black, very thick, mark=square*, mark size=1pt] coordinates {
  (-0.155,0.07) (-0.145,0.11) (-0.135,0.14) (-0.125,0.17) (-0.115,0.21) (-0.105,0.26) (-0.095,0.28) (-0.085,0.36) (-0.075,0.44) (-0.065,0.57) (-0.055,0.83) (-0.045,1.15) (-0.035,1.83) (-0.025,3.62) (-0.015,9.73) (-0.005,28.94) (0.005,29.95) (0.015,10.55) (0.025,4.01) (0.035,1.92) (0.045,1.23) (0.055,0.83) (0.065,0.61) (0.075,0.48) (0.085,0.36) (0.095,0.34) (0.105,0.31) (0.115,0.24) (0.125,0.18) (0.135,0.14) (0.145,0.12)
};
\addlegendentry{Pretrained}
\addplot[red!60!black, very thick, mark=triangle*, mark size=1.2pt] coordinates {
  (-0.155,0.01) (-0.145,0.01) (-0.135,0.03) (-0.125,0.04) (-0.115,0.05) (-0.105,0.07) (-0.095,0.11) (-0.085,0.18) (-0.075,0.25) (-0.065,0.38) (-0.055,0.58) (-0.045,0.76) (-0.035,1.15) (-0.025,2.51) (-0.015,8.53) (-0.005,35.79) (0.005,33.15) (0.015,10.21) (0.025,3.09) (0.035,1.15) (0.045,0.61) (0.055,0.39) (0.065,0.27) (0.075,0.16) (0.085,0.11) (0.095,0.09) (0.105,0.09) (0.115,0.09) (0.125,0.06) (0.135,0.04) (0.145,0.04)
};
\addlegendentry{RLVR}
\end{axis}

\begin{axis}[
    name=plotB,
    at={(plotA.right of north east)}, anchor=left of north west,
    xshift=0.4cm,
    width=8.7cm, height=5.3cm,
    xlabel={\footnotesize bond-angle error (deg) \scriptsize ($\Delta\theta =$ post $-$ pre)},
    ylabel={\footnotesize \% of angles},
    xmin=-15.5, xmax=15.5, ymin=0, ymax=20,
    xtick={-15,-10,-5,0,5,10,15},
    ytick={0,5,10,15,20},
    tick label style={font=\scriptsize},
    label style={font=\scriptsize},
    enlarge x limits=false,
    axis on top,
    legend style={font=\tiny, at={(0.97,0.95)}, anchor=north east, draw=gray!50},
    legend cell align=left,
]
\addplot[blue!55!black, very thick, mark=square*, mark size=1pt] coordinates {
  (-15.5,0.07) (-14.5,0.10) (-13.5,0.11) (-12.5,0.14) (-11.5,0.19) (-10.5,0.28) (-9.5,0.39) (-8.5,0.54) (-7.5,0.80) (-6.5,1.27) (-5.5,2.10) (-4.5,3.37) (-3.5,5.36) (-2.5,8.35) (-1.5,11.36) (-0.5,15.09) (0.5,15.17) (1.5,11.54) (2.5,8.58) (3.5,5.58) (4.5,3.49) (5.5,2.15) (6.5,1.30) (7.5,0.84) (8.5,0.56) (9.5,0.40) (10.5,0.27) (11.5,0.21) (12.5,0.16) (13.5,0.12) (14.5,0.11)
};
\addlegendentry{Pretrained}
\addplot[red!60!black, very thick, mark=triangle*, mark size=1.2pt] coordinates {
  (-15.5,0.01) (-14.5,0.01) (-13.5,0.02) (-12.5,0.03) (-11.5,0.05) (-10.5,0.07) (-9.5,0.10) (-8.5,0.17) (-7.5,0.31) (-6.5,0.60) (-5.5,1.24) (-4.5,2.45) (-3.5,4.70) (-2.5,8.55) (-1.5,13.49) (-0.5,18.19) (0.5,18.20) (1.5,13.66) (2.5,8.76) (3.5,4.67) (4.5,2.31) (5.5,1.13) (6.5,0.56) (7.5,0.28) (8.5,0.16) (9.5,0.10) (10.5,0.05) (11.5,0.04) (12.5,0.03) (13.5,0.03) (14.5,0.02)
};
\addlegendentry{RLVR}
\end{axis}
\end{tikzpicture}
\caption{\textbf{Per-bond and per-angle structural error introduced by the GFN2-xTB relaxation}. 
  Per-bond and per-angle errors. \emph{Left:} bond-length displacement $\Delta r =$ post $-$ pre, binned at
  $0.01\,\angstrom$. \emph{Right:} bond-angle displacement $\Delta\theta =$ post $-$ pre, binned at
  $1^{\circ}$. Both distributions are sharply peaked at zero (median
  $|\Delta r| = 0.008\,/\,0.006\,\angstrom$ and $|\Delta\theta| = 1.9^{\circ}\,/\,1.5^{\circ}$ for
  pretrained\,/\,RLVR).}
\label{fig:bond-angle-errors}
\end{figure*}
\label{sec:structural}

Since XTP-accepted molecules are topology-preserving, we
can compare each bond length and bond angle on the same
molecule. In Figure~\ref{fig:bond-angle-errors},
we show their distributions for unconditional (Btriple) generations
for the pretrained model and the RLVR model.

Both distributions are sharply peaked at zero. Median bond-length error is
$0.008\,\angstrom$ (pretrained model), and median
bond-angle error is $1.9^{\circ}$. 
We read this not as a precision claim but as supporting evidence that \xvr{} works well.
The XTP-accepted molecules sit close to the GFN2-xTB minimum, so relaxation
barely moves them. In other words, ADT works well 
to predict heavy-atom positions without relaxation.

RLVR sharpens this geometry. The XTP-accepted RLVR molecules keep the same
sharply-peaked shape, and the medians shrink: $|\Delta r|$ from $0.008$ to $0.006\,\angstrom$ and
$|\Delta\theta|$ from $1.9^{\circ}$ to $1.5^{\circ}$. This follows from the reward, which pays for low
relaxation strain. Pushing a geometry onto the GFN2-xTB minimum is exactly what shrinks the pre/post
displacement, so the fine per-bond geometry improves alongside topology preservation, not at its
expense.

\bigskip

\subsection{Scaling problem and the Inverse-Kinematics Transformer}
\label{sec:ikt}

We now turn to a separate issue: the scaling problem of generating long molecules.
In principle, we might expect \adt{} trained on short molecules to generate even long ones,
because the grammar to build a correct molecule can be short-range.
In practice, however, errors accumulate, as follows.

\begin{figure*}[t!]
\centering
\includegraphics[width=0.72\linewidth]{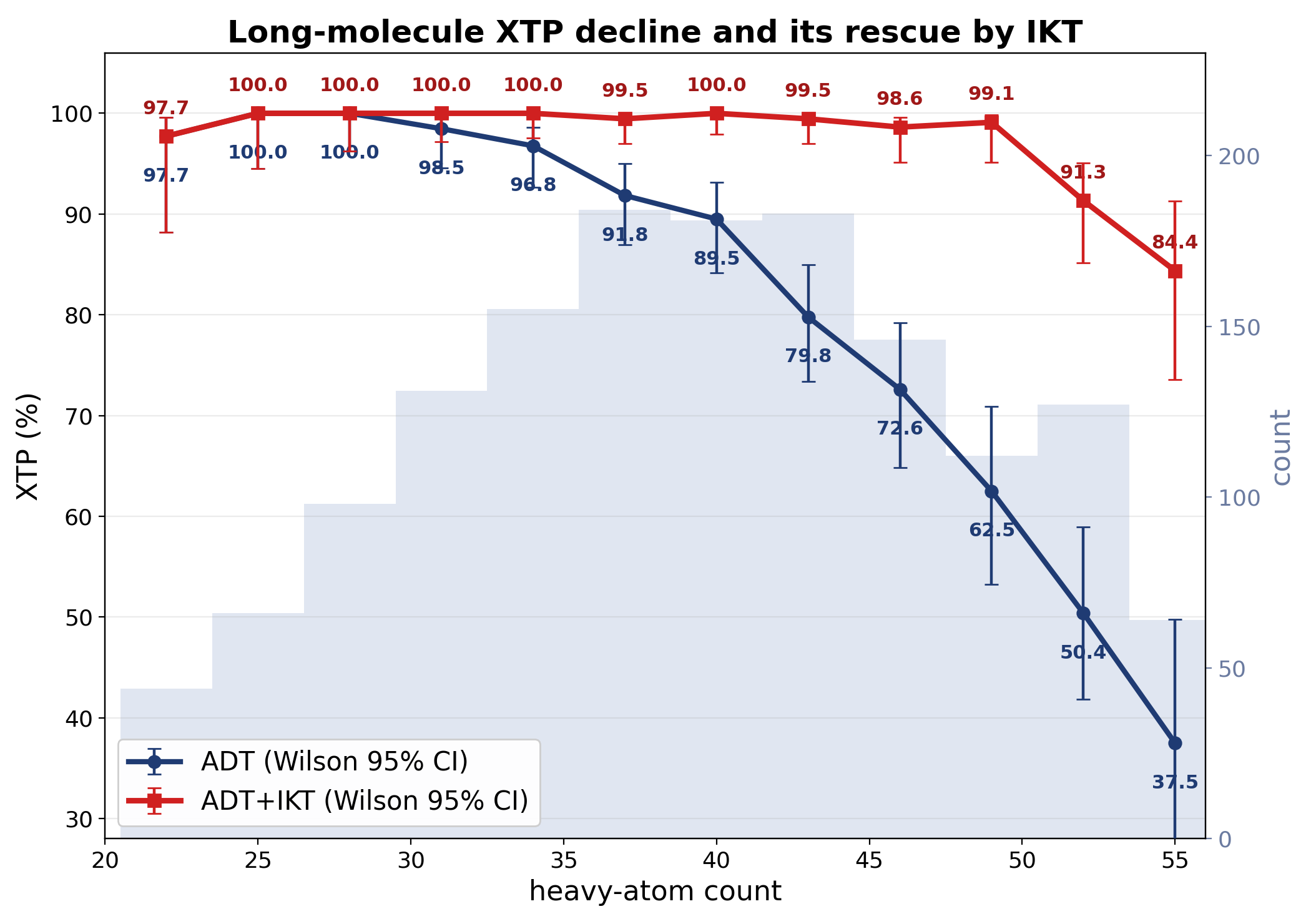}
\caption{\textbf{The Inverse-Kinematics Transformer repairs the
  large-molecule \xvr{} decline.} \xvr{} versus heavy-atom
  count for \adt{} (blue) and \adt{}$+$\ikt{} (red), with Wilson $95\%$ confidence
  intervals. Generation is biased toward long molecules to populate
  the 35--55-atom regime; the shaded histogram (right axis) is the
  per-bin sample count that sets each interval. \adt{}$+$\ikt{} may try up to six \ikt{} corrections per molecule,
  so the two curves are not at equal compute. \adt{} alone falls
  from $\sim$100\% below 34~atoms to $38\%$ at 55.}
\label{fig:ikt}
\end{figure*}

\adt{} places one heavy atom per step, so a small placement error is committed at each step. These
errors accumulate along the autoregressive trajectory, and
for large molecules the declared topology increasingly fails to relax to a topology-preserving minimum.
To quantify this, we bias generation toward long molecules by suppressing the end-of-sequence token,
producing a size distribution centered near $40$ heavy atoms (right axis in Figure~\ref{fig:ikt}).
The blue curve of Figure~\ref{fig:ikt} shows the \xvr{} as a function of heavy-atom count:
it is near $100\%$ below $\sim$34 atoms and drops to $38\%$ at $55$ atoms. 

We introduce the Inverse-Kinematics Transformer (\ikt{}), which repairs this problem. 
\ikt{} is a bidirectional (BERT-like) transformer that reads the completed \adt{} token streams and emits corrections,
a torsion and a bend per bond (one per {\tt ADD} or {\tt LINK} step).
We can train \ikt{} with RL, without initial data.
We let \ikt{} generate several candidate corrections (up to six in Figure~\ref{fig:ikt}).
If a corrected candidate passes \xvr{}, we accept it. \adt{} alone is a single draw, so the comparison
in Figure~\ref{fig:ikt} is not at equal compute: \adt{}$+$\ikt{} may spend up to six correction attempts on a molecule.
\ikt{} is complementary to \adt{}. Thanks to the bidirectional transformer, \ikt{} 
sees all the token streams at once.

The red curve of Figure~\ref{fig:ikt} shows
that \ikt{} lifts the \xvr{} from $89.5\%$ to $100.0\%$ at $40$ atoms and from $38\%$ to $84\%$ at $55$, largely flattening the decline.
\ikt{}'s teachers span $35$--$55$ heavy atoms, 
so Figure~\ref{fig:ikt} shows that it repairs the accumulated error in the regime it was trained for,
but not that it generalizes beyond it. At 55 atoms, \xvr{} is only $\sim 84\%$.
Thus we conclude that \adt{}+\ikt{} is promising, but generating accurate long molecules may need a
step-by-step size climb rather than a single jump: the target size is raised in stages, and at each
stage \ikt{} corrects the freshly accumulated error before the next, so the model bootstraps toward
sizes it cannot reach directly. This parallels how large language models are extended to long contexts
by curriculum, trained on shorter sequences first and lengthened progressively.

\section{Conclusion}
\label{sec:discussion}
\label{sec:summary}

\adt{} is a fully-discrete autoregressive transformer that places atoms one at a time, using a
parent-anchored local coordinate frame and a 7-slot discrete token vocabulary (action, parent offset,
atomic number, log-spaced distance bin, three HEALPix direction bins). 

We trained two models. One is the model pretrained on GEOM-Drugs, the other is the
RLVR model refined with the verifiable GFN2-xTB reward. Trained with no additional data,
the RLVR model reaches an end-to-end molecule generation rate of $\sim 95\%$.
This replaces imitation with self-improvement against the rules of the domain.

We summarize the three properties of the system that determine its practical usability as a
3D-molecule proposer: the reliability of the geometries it produces, the diversity of the chemistry it
explores, and the speed of generation.

\begin{itemize}

\item \textbf{Reliability.} The molecules \xvr{} accepts barely move
  under xTB relaxation: pre/post bond-length and bond-angle displacements
  (Figure~\ref{fig:bond-angle-errors}) are small (median
  $\sim 0.006\,\angstrom{}$ and $\sim 1.5^{\circ}$ for the RLVR model; $0.008\,\angstrom{}$ and
  $1.9^{\circ}$ for the pretrained model), evidence that the XTP-accepted geometries are
  self-consistent with the GFN2-xTB reference, rather than a precision
  claim against experiment. RLVR keeps the distribution sharply peaked
  and in fact sharpens it further (\S\ref{sec:structural}).

\item \textbf{Diversity.} 
Within each scaffold's RDKit-read-back set ($N_{\mathrm{XTP}}^{\mathrm{smiles}}$),
  uniqueness is at least $96.2\%$ for six of the seven scaffolds (cyclohexane: $94.4\%$),
  novelty against the full GEOM-Drugs set is at least $99.5\%$,
  Bemis--Murcko scaffold uniqueness is $49$--$66\%$, and
  median pair Tanimoto similarity is $\sim 0.14$--$0.21$,
  comparable to random drug pairs in
  ChEMBL~\citep{chembl2019}.
  \adt{} does not collapse onto a narrow
  chemical sub-region; a single trained model produces
  $\sim 4{,}500$--$7{,}400$ chemically distinct cores per scaffold
  (Table~\ref{tab:divers30}).

\item \textbf{Speed.} Sample-by-sample (no batching) generation on
  a single RTX~4090 takes $\sim 20\,\mathrm{ms}$ per added atom and
  scales approximately linearly with atom count over the sizes tested ($\sim 0.6\,\mathrm{s}$ for a
  30-atom molecule). Diffusion-based 3D generators instead incur a fixed
  denoising-step cost (typically several hundred steps) multiplying an
  $\mathcal{O}(N^{2})$ all-atom attention at each step, so, although we have not timed them under identical conditions, \adt{} should scale more
  favorably than iterative diffusion sampling in practice.

\end{itemize}

Together these make \adt{} a fast generator for scaffold-constrained 3D generation, the setting
underlying fragment growing and scaffold hopping. 
Several caveats temper the results: training uses only one low-energy conformer per molecule,
borderline \ch{C-S}/\ch{N-S} bonds and rare phosphorus chemistry are not yet well learned, and \xvr{}
rewards xTB topology preservation only, so properties with no reward term (notably QED and $\log P$) drift from
the GEOM-Drugs reference. Adding a property or binding term to the verifiable reward would steer these
axes directly. For longer molecules, the Inverse-Kinematics Transformer
(\S\ref{sec:ikt}) corrects the accumulated placement error while preserving topology.

Thus we believe that \adt{}+\ikt{} presented here might be a step toward handling the
complicated conditional 3D generation required for practical problems.

\section*{Reproducibility and code availability}
\label{sec:repro}

The code, the \xvr{} evaluation tool and a step-by-step guide (\texttt{REPRODUCE.md}) are at
\url{https://github.com/tkotani/ADT} (release \texttt{v2.1}). The checkpoints (the paper's numbers come from
\texttt{rlvr\_E240direct.pt} and \texttt{epoch\_240.pt}), the scaffold frame caches and the GEOM-Drugs SMILES list are
on Zenodo (\url{https://doi.org/10.5281/zenodo.20635985}).

To check the results, tell an AI coding agent (e.g.\ Claude Code) on a Linux machine with a CUDA GPU
(RTX~4090 or better recommended): ``Read \texttt{README\_reproduce\_benzene.md} in
\url{https://github.com/tkotani/ADT} and do what it says.'' It regenerates the benzene row of
Table~\ref{tab:drugs30} from the released checkpoint and compares it with the paper.

The implementation was written with extensive assistance from Anthropic's Claude (Opus); the author designed
the algorithms, verified the implementation, ran the computations, and takes responsibility for all results
and conclusions.

\section*{Patent disclosure}

This paper describes one specific implementation of the broader
framework disclosed in Japanese patent applications
No.~2026-16495 (filed 2026-01-17) and No.~2026-65995.

\section*{Acknowledgments}

The author thanks Prof.~Isao Tanaka, Prof.~Kazunori Sato, 
Prof.~Masao Obata, and Mr.~Kohki Hirai for helpful discussions and continuing support.

\bibliographystyle{unsrtnat}
\bibliography{adt}

\appendix
\renewcommand\appendixname{Appendix}

\begin{center}
{\Large\bfseries Appendix}
\end{center}
\vspace{0.5em}

\section{Token format details}
\label{app:token}

\subsection{Log-spaced distance bin $r_b$}
\label{app:rbin}

A continuous bond distance $r$ (in \AA) is discretized into one of $B=200$
log-spaced grid points indexed by $r_b\in\{0,1,\dots,B-1\}$ (the distance-bin
token of Table~\ref{tab:toksample}), spanning $R_{\min}=0.80\,\angstrom$ to
$R_{\max}=2.50\,\angstrom$:
\begin{equation}
  \hat{r}(r_b) \;=\; R_{\min}\,
  \Bigl(\tfrac{R_{\max}}{R_{\min}}\Bigr)^{\!r_b/(B-1)},
  \qquad r_b\in\{0,\dots,B-1\},
  \label{eq:rbin}
\end{equation}
so $r_b{=}0$ maps to $R_{\min}$ and $r_b{=}B{-}1$ to $R_{\max}$. A distance $r$
is encoded by clamping to $[R_{\min},R_{\max}]$ and rounding to the nearest grid
point in $\log r$; detokenization reads off $\hat{r}(r_b)$ from
Eq.~\eqref{eq:rbin}.

\subsection{HEALPix-encoded direction $(h_0, h_1, h_2)$}
\label{app:hpix}

A unit direction $\hat{u}\in S^2$ is encoded by HEALPix~\citep{healpix2005},
which partitions the sphere into pixels of equal solid angle at any
chosen $N_{\mathrm{side}}$. We use $N_{\mathrm{side}}{=}16$, giving
$N_{\mathrm{pix}}{=}12\,N_{\mathrm{side}}^{2}{=}3072$ pixels, each
subtending a solid angle $\Omega_{\mathrm{pix}} = 4\pi/(12\,N_{\mathrm{side}}^{2})
= 4\pi/3072 \approx 4.09\times 10^{-3}\,\mathrm{sr} \approx 13.4\,\mathrm{deg}^{2}$,
i.e.\ a typical pixel ``edge'' of $\sim 3.7^{\circ}$. The single nested
pixel index $p\in\{0,\ldots,3071\}$ is factored into three slots
$(h_0,h_1,h_2)$ that ADT stores as separate tokens:
\begin{equation}
\begin{aligned}
  p &= 256\,h_0 + 16\,h_1 + h_2,\\
  & h_0\in\{0,\ldots,11\},\quad h_1,h_2\in\{0,\ldots,15\}.
\end{aligned}
  \label{eq:hpix}
\end{equation}
$h_0$ selects one of the $12$ equal-area base faces and $(h_1,h_2)$
the $16{\times}16$ nested sub-pixels within it --- a hierarchical
quadrisection in which all $3072$ pixels share the same solid angle
$\Omega_{\mathrm{pix}}$, the equal-area property for which we adopt
HEALPix. Encoding/decoding ($\hat{u}\!\leftrightarrow\!p$) uses
\texttt{healpy.vec2pix}/\texttt{pix2vec} in nested mode.

\subsection{Tree-local frame for \texttt{ANGLE}}
\label{app:frame}

For the \texttt{ANGLE} action (step~2 of the Btriple), a local frame
is built from the first three atoms. The direction slots $(h_0,h_1)$ are reused
to hold two angle bins $(\theta_c,\theta_f)$ relative to this frame, and $h_2$
is unused. Subsequent \texttt{ADD}/\texttt{LINK} actions use HEALPix of
Eq.~\eqref{eq:hpix} in the parent's local frame.

\section{Hydrogen completion (MLnH) and placement (MLHplacer)}
\label{app:hcomplete}

\adt{} generates heavy-atom skeletons, so before an xTB relaxation we must add the right number of
hydrogens in the right places; two machine-learned heads do this (Phase~1 of \S\ref{sec:xvr}).
\textbf{MLnH} predicts a per-heavy-atom hydrogen count $n_{\mathrm{H}}$ from the local 3D environment,
after which a $\pm1$\,H parity correction adjusts the least-well-determined atom so the total
electron count is even, consistent with a neutral closed-shell xTB calculation. A learned count is
necessary because inferring $n_{\mathrm{H}}$ from an RDKit valence model first requires a clean
Lewis/kekulized perception of the heavy-atom graph, which fails on the aromatic and hypervalent motifs
common in drug-like sets: supplying hydrogens from an RDKit valence model alone produces a closed-shell,
xTB-valid molecule for only ${\sim}25\%$ of one pretrained set, whereas the learned count with the
parity correction reaches ${\sim}99.7\%$. \textbf{MLHplacer} then positions the $n_{\mathrm{H}}$
hydrogens in 3D with a second learned head (with a VSEPR-geometry fallback if it fails). Both heads
correct rather than reject, so they leave the survivor count of \S\ref{sec:xvr} unchanged
($N_{\mathrm{noclash}}{=}N_{n\mathrm{H}}{=}N_{\mathrm{Hplacer}}$); their role is to make the
hydrogen-completed molecule survive the subsequent xTB relaxation.

\section{The two heavy-atom topologies}
\label{app:topology}

\paragraph{The ADT graph $\badt$.} The model does not only place atoms; it says what it is
bonding. Each \texttt{ADD} step names a parent, and each \texttt{LINK} step names a ring closure, so the token stream carries a graph outright (the atom set is fixed, so the graph is its edge
set):
\begin{equation}
  \badt \;=\; \underbrace{\bigl\{\,(\mathrm{parent}(k),\,k)\,\bigr\}_{k}}_{\text{the rooted tree } T}
  \;\cup\;
  \underbrace{\bigl\{\,(i,j)\ \text{emitted as \texttt{LINK}}\,\bigr\}}_{\text{ring closures}} .
  \label{eq:badt}
\end{equation}
$\badt$ has no bond orders and no coordinates. It is what the model declared it was building.
The symbol is the ordinary graph-theoretic $G$; the edge-set symbol $E$ is unavailable here, being the
energy.

\paragraph{The perceived graph $\bper{X}$.} Given heavy-atom positions $\{\mathbf{x}_i\}$ with
atomic numbers $\{Z_i\}$, atoms $i$ and $j$ are taken to be bonded iff
\begin{equation}
  \lVert \mathbf{x}_i - \mathbf{x}_j \rVert \;<\; 1.3\,\bigl(r_{\mathrm{cov}}(Z_i) + r_{\mathrm{cov}}(Z_j)\bigr),
  \label{eq:bond}
\end{equation}
with Cordero covalent radii $r_{\mathrm{cov}}$~\citep{cordero2008} (in \angstrom): H~$0.31$, C~$0.76$, N~$0.71$,
O~$0.66$, F~$0.57$, P~$1.07$, S~$1.05$, Cl~$1.02$, Br~$1.20$, I~$1.39$.
This yields the perceived graph $\bper{X}$ of any heavy-atom configuration $X$: the atoms are
its vertices, the bonded index pairs its edges.

The distance cutoff of Eq.~\eqref{eq:bond} is somewhat arbitrary, but on drug-like geometries it
reliably tells bonded from non-bonded pairs. Across the accepted molecules (writing
$r = d/1.3(r_i{+}r_j)$, so the cutoff is $r{=}1$), every bond has $r \le 0.99$ and every non-bonded
contact $r \ge 1.00$, a gap that never closes in any molecule. The cutoff being respected on the data,
rewarding $\bper{X}{=}\badt$ under RL enforces a genuine chemical topology, not an artifact of the rule.

\section{RLVR method and hyperparameters}
\label{app:rlvr}

This appendix details the RLVR of \S\ref{sec:drugs50}. Starting from
the pretrained policy $\pi_{\theta}$, we maximize a single objective by
policy-gradient (REINFORCE):
\begin{equation}
  \begin{split}
    \mathcal{L} \;=\;& \underbrace{\mathcal{L}_{\mathrm{RL}}}_{\text{validity reward}}
    \;+\; \underbrace{\textstyle\sum_{k}\lambda_{k}\,\mathcal{L}_{\mathrm{ctrl}}^{k}}_{\text{extra reward}}
    \;+\; \lambda_{\mathrm{t}}\,\mathrm{KL}\!\left(\pi_{\theta}\,\|\,\pi_{\mathrm{ref}}\right)^{2} \\
    &+\; \lambda_{s}\,\mathrm{KL}\!\left(P_{\mathrm{tgt}}\,\|\,\hat P_{\theta}\right)
    \;+\; \lambda_{c}\,\mathrm{KL}\!\left(f_{\mathrm{tgt}}\,\|\,\bar f_{\theta}\right),
  \end{split}
  \label{eq:rlvr-loss}
\end{equation}
where the REINFORCE term
\begin{equation}
  \mathcal{L}_{\mathrm{RL}} \;=\;
  -\,\mathbb{E}_{\tau\sim\pi_{\theta}}\!\Big[\big(r_{\tau}-r_{\mathrm{base}}\big)\,\log\pi_{\theta}(\tau)\Big]
  \label{eq:reinforce}
\end{equation}
takes its expectation over trajectories sampled from the policy: starting from
the empty prefix, actions are drawn one at a time,
$a_{t}\sim\pi_{\theta}(\cdot\mid s_{t})$, each conditioned on the state
$s_{t}=(a_{0},a_{1},\dots,a_{t-1})$ --- the actions chosen so far --- until the
\texttt{END} action closes the sequence $\tau=(a_{0},a_{1},\dots,\texttt{END})$,
one completed molecule. Its log-likelihood factorizes as
$\log\pi_{\theta}(\tau)=\sum_{t}\log\pi_{\theta}(a_{t}\mid s_{t})$. The
size-conditioned baseline $r_{\mathrm{base}}$ (an exponential moving average of the
reward within each heavy-atom count) reduces the gradient variance, and
$r_{\tau}\in[0,1]$ is
the verifiable reward on the completed $\tau$.
The reward is the physical \xvr{} criterion (\S\ref{sec:xvr}): a
GFN2-xTB relaxation must realize the graph $\badt$ the model declared. It is graded by relaxation
energy as detailed below.

\paragraph{Validity reward term.}
We reward \xvr{}, graded by the relaxation energy:  
\begin{equation}
  r_{\tau} \;=\;
  \begin{cases}
    0, & \text{not \xvr{}},\\[2pt]
    0.6 + 0.4\,\exp(-\varepsilon/T_{\varepsilon}), & \text{\xvr{}},
  \end{cases}
  \label{eq:reward}
\end{equation}
with $\varepsilon = (E_{\mathrm{Hprerx}} - E_{\mathrm{fullrx}})/N_{\mathrm{heavy}}$ the per-heavy-atom
energy released as the hydrogen-pre-relaxed structure descends to the free minimum.
$T_{\varepsilon} = 2.0$~kcal\,mol$^{-1}$ per atom. There is no partial credit: a molecule that relaxes
to a different graph scores $0$, so the reward gap of $0.6$ holds everywhere. 

\paragraph{Extra reward term.}
Beyond validity, the reward may carry a few optional control terms. Each has the
same REINFORCE form as Eq.~\ref{eq:reinforce}, but with a population quantity in
place of the reward:
\begin{equation}
  \mathcal{L}_{\mathrm{ctrl}}^{k} \;=\;
  -\,\mathbb{E}_{\tau\sim\pi_{\theta}}\!\Big[\big(q^{k}_{\tau}-q^{k}_{\mathrm{base}}\big)\,\log\pi_{\theta}(\tau)\Big],
  \label{eq:ctrl}
\end{equation}
where $q^{k}_{\tau}$ summarizes quantity $k$ on molecule $\tau$ and
$q^{k}_{\mathrm{base}}$ is its running mean. Because the baseline centers the advantage
($\sum_{\tau}(q^{k}_{\tau}-q^{k}_{\mathrm{base}})=0$), the term redistributes gradient toward the
wanted side of $q^{k}$ with no net push; its weight $\lambda_k$ is a signed PID gain (positive for a
floor, negative for a ceiling) that acts only while $q^{k}$ breaches its target. The steerable
quantities are either instantly evaluable per molecule ($\log P$, molecular size, element composition)
or accumulative, such as the Bemis--Murcko scaffold, whose rarity ($1/\mathrm{count}$) is defined only
over a rolling window. This run uses a scaffold-diversity floor (uniqueness $\geq 0.96$); any RDKit
descriptor can be steered the same way, a data-free knob that leaves the validity reward untouched.

\begin{table*}[htb]
  \centering\small
  \setlength{\tabcolsep}{5pt}
  \begin{tabular}{lll}
    \toprule
    symbol & controls & status \\
    \midrule
    $\lambda_{\mathrm{t}}$     & trust-region weight        & fixed, $0.1$ \\
    $\lambda_s$                & size-KL weight             & fixed, $1.0$ \\
    $\lambda_c$                & composition-KL weight      & adaptive: PID (base $2.0$) \\
    $\lambda_{\mathrm{scaf}}$  & scaffold-floor weight      & adaptive: signed PID, floor $0.96$ \\
    $\lambda_{\log P}$         & $\log P$-ceiling weight     & adaptive: signed PID, ceiling $2.9$ \\
    $(\mu,\sigma)$             & size virtual target        & stepped $\to$ realized $(25.4,5.5)$ \\
    $f_{\mathrm{tgt}}$         & composition virtual target & stepped $\to$ realized GEOM marginal \\
    $\pi_{\mathrm{ref}}$       & trust-region reference     & re-anchored each ratchet \\
    ratchet   & reference/target clock     & every $100$ steps \\
    $N_{\max}$  & size ceiling             & $56$ heavy atoms \\
    $T_{\varepsilon}$  & reward strain temperature  & fixed, $2.0$~kcal\,mol$^{-1}$/atom \\
    \bottomrule
  \end{tabular}
  \caption{RLVR controls of Eq.~\ref{eq:rlvr-loss}, plus the reward strain temperature
    $T_{\varepsilon}$ of Eq.~\ref{eq:reward}, and their status. Loss weights
    are fixed or online-adapted by a PID; the size and composition targets are
    virtual --- stepped so the realized distribution lands on GEOM,
    not the target itself; a reward control acts through a floor/ceiling on a
    realized quantity, its signed PID gain switching direction for a floor
    vs.\ a ceiling. The reported run uses every row except
    $\lambda_{\log P}$, an optional control shown for illustration.}
  \label{tab:controls}
\end{table*}

\paragraph{KL terms.}
The three KL terms share the ratchet clock: every $100$ steps the reference
each is measured against is refreshed --- $\pi_{\mathrm{ref}}$ is re-anchored to the
current policy, and the size and composition virtual targets are stepped (below).
\emph{(i) Trust region.} The squared-KL to $\pi_{\mathrm{ref}}$ keeps each
update local and, with re-anchoring, lets the trust region walk uphill across
plateaus; squaring is essential --- a signed/linear KL diverges once the reward
saturates.
\emph{(ii) Size.} $P_{\mathrm{tgt}}$ is a censored Gaussian $\mathcal{N}(\mu,\sigma)$
over heavy-atom count and $\hat P_{\theta}$ the model's censored at-risk
distribution (the KL summed over counts).
\emph{(iii) Composition.} $f_{\mathrm{tgt}}/\bar f_{\theta}$ are the target/model
element marginals.
Of the three weights, the trust-region $\lambda_{\mathrm{t}}$ and size $\lambda_{s}$
are fixed ($0.1$ and $1.0$), while the composition $\lambda_{c}$ is held by a PID
thermostat.

\paragraph{Virtual targets.}
The size and composition targets are virtual: we do not fix them at the
GEOM-Drugs values but step them so the realized distribution equilibrates
there. A fixed target cannot cancel a systematic pull from the reward --- at any
finite weight the equilibrium sits offset by the reward's bias --- so at each
ratchet the size moment $(\mu,\sigma)$ of $P_{\mathrm{tgt}}$ and the element target
$f_{\mathrm{tgt}}$ are nudged (a capped $\pm$step within a tolerance band) in the
direction that carries the realized moment / marginal toward the desired GEOM
value, gated on \xvr{} so the loop advances only while validity holds. The size
target is two numbers $(\mu,\sigma)$ (the KL is summed over heavy-atom counts); the
composition target is a per-element marginal. The realized size $(25.4,5.5)$ and
element marginal (Table~\ref{tab:drugs30honest}) thus land on GEOM even under
the reward's size/composition pull; the scaffold floor and $\log P$ ceiling of
Eq.~\ref{eq:ctrl}, by contrast, act on realized quantities directly and need no
virtual target. Table~\ref{tab:controls} lists every control and its status.

\paragraph{Effect on the output distribution.}
These controls shape the generated distribution, not only the loss.
After RLVR the XTP-accepted output retains the GEOM-Drugs heavy-atom-count
moment (mean $25.1$, $\sigma \approx 5.5$ across scaffolds;
Table~\ref{tab:drugs30}) and --- held by the composition term --- the
GEOM-Drugs element marginal (Table~\ref{tab:drugs30honest}), so validity is amplified without shifting
molecular size or atom-type makeup. Only properties with no reward or
control term in this run, notably QED and Crippen $\log P$, drift from the
reference (Table~\ref{tab:divers30}); either can be pinned back by such a
control term (Eq.~\ref{eq:ctrl}) at a small diversity cost.
\section{Training curves}
\label{app:training}

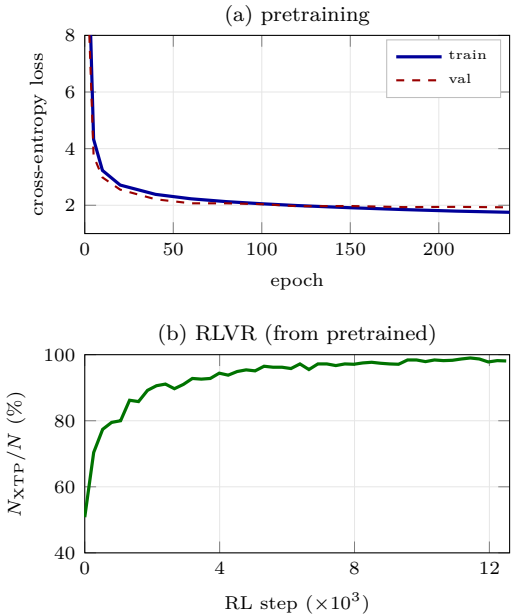
\begin{figure}[t!]
\centering
\begin{tikzpicture}
\begin{groupplot}[
    group style={group size=1 by 2, vertical sep=1.6cm},
    width=7.2cm, height=4.2cm,
    tick label style={font=\scriptsize},
    label style={font=\scriptsize},
    legend style={font=\tiny, draw=gray!50},
    legend cell align=left,
    grid=major, grid style={gray!20, very thin},
]
\nextgroupplot[
    title={\footnotesize (a) pretraining},
    title style={yshift=-0.6em},
    xlabel={epoch}, ylabel={cross-entropy loss},
    xmin=0, xmax=240, ymin=1, ymax=8,
]
\addplot[blue!60!black, very thick] coordinates {
  (1,13.1906) (5,4.3287) (10,3.2300) (20,2.7134) (40,2.3847) (60,2.2283) (80,2.1283)
  (100,2.0531) (120,1.9897) (150,1.9129) (180,1.8480) (210,1.7927) (240,1.7518)
};
\addlegendentry{train}
\addplot[red!60!black, thick, dashed] coordinates {
  (1,10.8344) (5,3.7388) (10,2.9837) (20,2.5522) (40,2.2121) (60,2.0689) (80,2.0689)
  (100,2.0396) (120,1.9671) (150,1.9730) (180,1.9350) (210,1.9433) (240,1.9262)
};
\addlegendentry{val}
\nextgroupplot[
    title={\footnotesize (b) RLVR (from pretrained)},
    title style={yshift=-0.6em},
    xlabel={RL step ($\times 10^{3}$)}, ylabel={$N_{\mathrm{XTP}}/N$ (\%)},
    xmin=0, xmax=12.6, ymin=40, ymax=100,
    xtick={0,4,8,12}, ytick={40,60,80,100},
]
\addplot[green!45!black, very thick] coordinates {
  (0,50.8) (0.265,70.4) (0.531,77.4) (0.797,79.5) (1.063,80.0) (1.329,86.2) (1.595,85.8) (1.861,89.2)
  (2.127,90.6) (2.393,91.1) (2.659,89.7) (2.925,91.0) (3.191,92.8) (3.457,92.6) (3.723,92.8) (3.989,94.4)
  (4.254,93.8) (4.52,94.9) (4.786,95.4) (5.052,95.1) (5.318,96.5) (5.584,96.2) (5.85,96.2) (6.116,95.8)
  (6.382,97.2) (6.648,95.5) (6.914,97.2) (7.18,97.2) (7.446,96.7) (7.712,97.2) (7.978,97.1) (8.244,97.5)
  (8.509,97.7) (8.775,97.4) (9.041,97.2) (9.307,97.1) (9.573,98.4) (9.839,98.4) (10.105,97.9) (10.371,98.4)
  (10.637,98.2) (10.903,98.3) (11.169,98.7) (11.435,99.0) (11.701,98.7) (11.967,97.8) (12.233,98.2) (12.499,98.1)
};
\end{groupplot}
\end{tikzpicture}
\caption{\textbf{Training curves.}
  \textbf{(a)}~Pretraining: training and validation cross-entropy loss versus epoch for the Drugs
  model (\S\ref{sec:drugs30}).
  \textbf{(b)}~Data-free RLVR from the pretrained E240 model: batch topology-preservation rate
  $N_{\mathrm{XTP}}/N$ versus RL step ($\times 10^{3}$), a $25$-step moving average. It reaches
  ${\sim}98\%$ at RL step ${\sim}10{,}000$ and holds. Cost: ${\sim}10{,}000$ steps ${\approx}\,2.3$
  days on one RTX~4090 with $16$ CPU cores (${\sim}20$\,s/step, dominated by the three GFN2-xTB
  relaxations per molecule).}
\label{fig:training-curves}
\end{figure}

Figure~\ref{fig:training-curves}(a) shows training and validation
cross-entropy loss versus epoch for the pretrained Drugs model
(\S\ref{sec:drugs30}), trained from scratch to E240.
We train with dropout, which keeps the validation loss close to the training loss
throughout; that near-agreement is the indicator we watch --- it signals the model is generalizing
rather than memorizing --- and we stop pretraining at E240 while the two still track.

Figure~\ref{fig:training-curves}(b) is the headline result: data-free RLVR converges, taking
$N_{\mathrm{XTP}}/N$ from ${\sim}51\%$ to ${\sim}98\%$ and holding there (details in the caption).

The RLVR stage (\S\ref{sec:drugs50}) uses a batch of $48$ molecules per step with $16$ parallel
xTB workers; the reward is \xvr{} with per-atom strain shaping (Eq.~\ref{eq:reward}) and the controls
of Table~\ref{tab:controls} (Appendix~\ref{app:rlvr}). Exact launch commands are logged for
reproducibility.

\section{Example generated molecules}
\label{app:examples}

We list two representative XTP-accepted generations per scaffold for the
RLVR model
(Table~\ref{tab:gen-rlvr}; from a run with the earlier RLVR checkpoint, see its caption). All SMILES are canonical
non-isomeric.
$\Delta E_{\text{xTB}}{=}E_{\mathrm{Hprerx}}{-}E_{\mathrm{fullrx}}$ is the per-molecule
GFN2-xTB relaxation strain (kcal/mol; positive); $\text{RMSD}_{h}$ is the heavy-atom RMSD between pre- and
post-xTB positions ($\angstrom$; the subscript $h$ denotes heavy atoms,
i.e.\ hydrogens excluded).

\clearpage
\begin{table*}[t!]
\centering\footnotesize
\setlength{\tabcolsep}{4pt}
\begin{tabular}{l@{\hskip 6pt}p{8.4cm}rr}
\toprule
\textbf{Scaffold} & \textbf{Canonical SMILES} & $\Delta E_{\text{xTB}}$\,(kcal/mol) & $\text{RMSD}_{h}$\,($\angstrom$) \\
\midrule
\multicolumn{4}{@{}l}{\textbf{(a) Typical}\ \footnotesize---\ two per scaffold, strain nearest that scaffold's median} \\[1pt]
benzene     & \smi{CCOC(=O)COc1ccc(Sc2ccc(Oc3ccccc3)cn2)nc1} & $9.6$ & $0.14$ \\
            & \smi{O=C(Nc1ccc(Cl)c(O)c1)c1ccc(C=NNc2ccccc2)nn1} & $9.6$ & $0.45$ \\
pyridine    & \smi{c1ccc(-c2ccc(Sc3cccnc3)cn2)cc1} & $7.7$ & $0.51$ \\
            & \smi{CCOc1ccc(Sc2ccc(OC(=O)c3ccncc3)cn2)cc1} & $7.7$ & $0.05$ \\
pyrimidine  & \smi{COc1ccnnc1-c1cc(SCC(=O)Nc2ccccc2)ncn1} & $9.1$ & $0.12$ \\
            & \smi{CCCOC(=O)c1cc(Sc2nc(N)c3ccccc3n2)ccc1F} & $9.1$ & $0.64$ \\
pyrazine    & \smi{C=COC(=O)COc1ccc(CSc2cnccn2)cc1} & $10.2$ & $0.12$ \\
            & \smi{O=C(C=Cc1ccc(CO)nc1)Nc1cnccn1} & $10.2$ & $0.21$ \\
furan       & \smi{COc1cc(OCC(=O)O)ccc1CC=NNC(=O)c1ccco1} & $8.4$ & $0.12$ \\
            & \smi{Cc1ccc(CNC(=O)c2ccc(CSc3cccnn3)o2)cn1} & $8.4$ & $1.10$ \\
thiophene   & \smi{CNC(=O)Oc1ccc(COC(=O)NCCc2cccs2)cc1Cl} & $8.4$ & $0.19$ \\
            & \smi{CCCOC(=O)c1cc(F)cc(OCc2ccc(C(=O)Nc3cccs3)cc2F)c1} & $8.4$ & $0.30$ \\
cyclohexane & \smi{CCCCSCCc1ccc(OC(=O)N2C3CCC2CC3)cc1F} & $7.9$ & $0.20$ \\
            & \smi{CC(F)CCOc1cccc(Sc2ccc(OC(=O)NC3CCCCC3)cn2)c1} & $7.9$ & $0.12$ \\
\midrule
\multicolumn{4}{@{}l}{\textbf{(b) No all-neutral Lewis structure, but a real motif}\ \footnotesize---\ an S$^+$ ylide or an N-oxide} \\[1pt]
thiophenium ylide & \smi{[O-]C(NN=Cc1ccccc1)=C1C=CC=[S+]1} & $1.4$ & $0.08$ \\
$N$-oxide\,/\,azinium & \smi{COc1ccc(-c2cn([O-])cc(C)[n+]2=O)cc1} & $5.2$ & $0.15$ \\
\midrule
\multicolumn{4}{@{}l}{\textbf{(c) Structures unlikely to be real}\ \footnotesize---\ xTB accepts them, GEOM-Drugs never does} \\[1pt]
carbanion\,/\,iminium & \smi{CCCCCCCOC(=O)c1ccc(-[n+]2[c-]c3c(ncn3C)nc2C)cc1} & $4.7$ & $0.13$ \\
carbanion\,/\,oxocarbenium & \smi{C=C([CH2-])OC(=O)c1cc(OC(=O)C2=C=CC=[O+]2)ccc1Cl} & $4.3$ & $0.12$ \\
carbocation & \smi{C=C([CH2+])Oc1cc(OC(=O)c2[c-]ccnc2)ccc1Br} & $5.4$ & $0.08$ \\
in-ring cumulene & \smi{CCCCCOC(=O)c1cc(OCCCN2C=C=CC=C2)ccc1F} & $10.5$ & $0.28$ \\
aryne (in-ring alkyne) & \smi{Cc1ccc(Cn2cnc(Oc3cc\#ccc3)c2)o1} & $16.4$ & $0.38$ \\
\bottomrule
\end{tabular}
\caption{\textbf{Representative generations from the RLVR model.}
  These examples come from a generation run with the earlier RLVR checkpoint used in version~1 of this
  preprint; the statistics of Tables~\ref{tab:drugs30}--\ref{tab:drugs30honest} are from the released checkpoint.
  All rows are XTP-accepted and
  are read as valid molecules by RDKit. SMILES are canonical non-isomeric, read off the relaxed
  geometry. $\Delta E_{\mathrm{xTB}}{=}E_{\mathrm{Hprerx}}{-}E_{\mathrm{fullrx}}$ is the per-molecule
  relaxation strain (kcal\,mol$^{-1}$) to the free minimum of the \emph{declared} topology, not a
  deviation from a global ground state; $\mathrm{RMSD}_h$ is the heavy-atom RMSD between the generated
  and relaxed geometry (\AA).
  (a) typical drug-like rows (QED${\geq}0.4$, $18$--$32$ heavy atoms, all
  formally neutral), two per scaffold with $\Delta E_{\mathrm{xTB}}$ nearest that scaffold's median in that run, so
  typical rather than selected.
  (b) charge-separated but real motifs: an S$^+$ ylide and an N-oxide.
  (c) structures unlikely to be real: off-sulfur carbanion/carbocation pairs and strained rings
  (an in-ring cumulene, an aryne). Blocks (b) and (c) are quantified and discussed in \S\ref{sec:drugs50}.}
\label{tab:gen-rlvr}
\end{table*}

\end{document}